\documentstyle[aps,preprint,tighten,floats,psfig]{revtex}
\begin{document}
\draft

\title{Predictive Ability of QCD Sum Rules for Excited Baryons}
\author{Frank X. Lee~$^{a,b}$ and Xinyu Liu~$^a$}
\address{Center for Nuclear Studies, Department of Physics,
$^a$~The George Washington University,  Washington, DC 20052, USA \\
$^b$~Jefferson Lab, 12000 Jefferson Avenue, Newport News, VA 23606, USA}

\maketitle
           
\begin{abstract}
The masses of octet baryons are calculated by the method of QCD sum rules.
Using generalized interpolating fields,
three independent sets of QCD sum rules are derived which allow the extraction of
low-lying N* states with spin-parity 1/2+, 1/2- and 3/2- in both the non-strange and
strange channels.
The predictive ability of the sum rules is examined by a
Monte-Carlo based analysis procedure in which the three phenomenological
parameters (mass, coupling, threshold) are treated as free parameters
simultaneously.
Realistic uncertainties in these parameters are obtained by simultaneously
exploring all uncertainties in the QCD input parameters.
Those sum rules with good predictive power are identified and their predictions are 
compared with experiment where available.
\end{abstract}
\vspace{1cm}

\section{Introduction}
\label{intro}

     The QCD sum rules method~\cite{SVZ79} is a time-honored approach in
revealing a direct connection between hadron phenomenology and
QCD vacuum structure via a few universal parameters called vacuum condensates
(vacuum expectation values of QCD local operators).
The most important are the quark condensate,
the mixed condensate and the gluon condensate.
The method has been successfully applied to a variety of
problems to gain a non-perturbative understanding
into the structure of hadrons (for a review
of the early work, see Ref.~\cite{RRY85}), and continues to be
an active field~\cite{archive}.

The use of QCD sum rules to study baryon masses started not long after the method was
introduced.
Early works include Ref.~\cite{Ioffe81,Bely82,Bely83,RRY82,Chung84,RRY85,Derek90,Derek96}
in which a variety of baryons were studied in the approach.
The success of this somewhat unconventional method
in predicting hadron properties has historically given us a unique perspective
on hadron structure, compared to other conventional views,
like the phenomenologically successful constituent quark model.

In recent years, there is increasing desire to understand the structure of
baryons directly from QCD, the fundamental theory of the strong interaction.
In particular, increasing focus is put on N* physics, the study of excited states
baryons. After all, most of the observed baryon spectrum under about 2 GeV
can be considered as some excitation of the ground-state nucleon.
On the experimental side, the field is fueled by the high-quality data
coming from JLab and other accelerators.
QCD is very difficult to solve in the low-energy regime.
While lattice QCD, which solves QCD with controlled systematic
errors on a discrete space-time lattice by numerical simulations,
holds the key to ultimately unravel hadron structure,
its results are still hard to obtain.
One the other hand, the QCD sum rule method can serve as a complement
to lattice QCD based on the qualifications that it has close contact with
the basic premise of QCD, has minimal model dependence with well-understood
limitations inherent in the operator product expansion (OPE).
One advantage of the method is that it provides a physically transparent
view of how hadrons arise from QCD via the vacuum condensates.
One can trace back, term by term, what operators and how these operators conspire
to give rise to the mass of a baryon, and mass splittings among baryons.
An example is the so-called Ioffe formula~\cite{Ioffe81},
$M_N\simeq (-8\pi^2<\bar{q}q>)^{1/3}=0.97$ GeV,
which is an over-simplified result by keeping only the leading order
contributions in the nucleon QCD sum rules.
It nonetheless gives a nucleon mass not too far from the observed value and
a direct link to the quark condensate,
the order parameter of spontaneous chiral symmetry breaking of QCD.
In general, the reliability of such predictions should be checked by
including all terms and by careful analysis.
But the insight is valuable in understanding hadron structure.

To make the QCD sum rules method a useful tool in probing N* structure,
a key question is how much predictive power the method has.
Within the limitations of the method,
the question is sum-rule dependent because all QCD sum rules are not equal,
as the examples will show in this work.
It depends on the type of correlation functions and the tensor structure
from which a sum rule is constructed because the OPE usually has
different convergence properties at each structure.
The main goal of this work is to examine the
predictive ability of the QCD sum rule approach in its standard implementation,
using the octet baryons as examples.
A similar study has been done for the decuplet baryons~\cite{Lee98}.
In our view, such a study is useful in a number of ways.

First,
we find the existing works on these baryons are incomplete and sometimes
inconsistent in a number of areas, such as the interpolating fields,
number of terms included in the OPE,
the anomalous dimension corrections, the continuum model,
and the treatment of strange baryons.
A perusal of these works sometimes reveals discrepancies among the QCD sum rules
where comparisons are possible.
Therefore, before we move to the issue of predictive power, we need a set of
sum rules as complete as possible.
To this end, we first derive a complete set of QCD sum rules in order to
independently verify and expand upon existing results.
We use generalized interpolating fields and three types of correlation functions among them. We study not only the nucleon, but also  the strange members of the octet family.  Some of the sum rules in this work are new.
The result is a complete, updated set of QCD sum rules under one roof
for octet baryons in the standard implementation of the approach
(11 sum rules for each of the $N$, $\Sigma$, $\Xi$ and $\Lambda$ channels)
which encompass the existing sum rules as special cases.
We consistently include operators up to dimension 8,
first order strange quark mass corrections,
flavor symmetry breaking of the strange quark condensates,
anomalous dimension corrections,
possible factorization violation of the four-quark condensate,
and a pole plus continuum model.

Second, it concerns how to extract a physical quantity from sum rules
and how reliable it is.
Usually there is considerable freedom in how the analysis is done
regarding the Borel mass window, the modeling of the continuum,
and the treatment of the fit parameters, each of which has some impact on the
outcome.
In the past, often an error is assigned to a quantity
without the support of rigorous error analysis.
A quantitative tool is needed to treat these degrees of freedom as complete as
possible.  This was first addressed by Leinweber~\cite{Derek96} who introduced
a Monte-Carlo based procedure that incorporates all uncertainties in the
QCD input parameters simultaneously, and maps them into
uncertainties in the phenomenological parameters, with careful regard to
OPE convergence and ground state dominance.
We will apply this procedure in the current work.

Third, the study will aid the application of the method to
matrix elements of N* (three-point functions),
such as magnetic moments, transition moments, axial charges,
tensor charges, etc., because they utilize the two-point function as normalization.

The paper is organized as follows. In section~\ref{method},
the interpolating fields are introduced, followed by a complete listing of
all the derived sum rules divided into three groups.
Section~\ref{analysis} discusses the analysis procedure and all the results.
Section~\ref{con} gives the concluding remarks.

\section{Method}
\label{method}

The basic idea behind the QCD sum rules approach is the ansatz of duality.  
That is, it is possible
to simultaneously describe a hadron as quarks propagating in the QCD
vacuum, and as a phenomenological field with the appropriate quantum numbers.
In this way, a connection can be established between a
description in terms of hadronic degrees of freedom
and one based on the underlying quark and gluon degrees of freedom governed
by QCD.

Hadron masses can be extracted from the time-ordered two-point correlation function 
in the QCD vacuum
\begin{equation}
\Pi(p)=i\int d^4x\; e^{ip\cdot x}\langle 0\,|\,
T\{\;\eta(x)\, \bar{\eta}(0)\;\}\,|\,0\rangle,
\label{cf2pt}
\end{equation}
where $\eta$ is the interpolating field (or hadron current)
with the quantum numbers of the baryon under consideration.

Assuming SU(2) isospin symmetry in the $u$ and $d$ quarks, we consider 
the most general current of spin 1/2 and isospin 1/2 for the nucleon:
\begin{equation}
\eta^{ N}_{1/2}(x)=
-2\;\left[\,\eta^{ N}_1(x) 
+ \beta\; \eta^{ N}_2(x)\,\right],
\label{nucleon_beta}
\end{equation}
where
\begin{equation}
\eta^{ N}_1(x)=\epsilon^{abc}
\left(u^{aT}(x)C\gamma_5 d^b(x)\right)u^c(x),
\end{equation}
and 
\begin{equation}
\eta^{ N}_2(x)=\epsilon^{abc}
\left(u^{aT}(x)C d^b(x)\right)\gamma_5 u^c(x).
\end{equation}
Here $C$ is the charge conjugation operator, the superscript $T$ means transpose,
and $\epsilon_{abc}$ makes it color-singlet.
The real parameter $\beta$ allows for the mixture of the two independent currents.
The choice advocated by Ioffe~\cite{Ioffe81} and often used in QCD sum rules studies
corresponds to $\beta=-1$.
In this work, the freedom to vary $\beta$ is exploited to achieve maximal 
overlap with the state in question for a particular sum rule.
It proves an effective tool to saturate a sum rule either by positive- or by negative-parity states. 

In the following we write down the interpolating fields 
for the other members of the octet, omitting the explicit $x$ dependence.
For $\Sigma$ we consider
\begin{equation}
\eta^{ \Sigma}=-2\,\epsilon^{abc}\left[
(u^{aT}C\gamma_5 s^b)u^c
+\beta(u^{aT}Cs^b)\gamma_5 u^c \right],
\end{equation}
and for $\Xi$
\begin{equation}
\eta^{ \Xi}=-2\,\epsilon^{abc}\left[
(s^{aT}C\gamma_5 u^b)s^c
+\beta (s^{aT}Cu^b)\gamma_5 s^c \right].
\end{equation}
For the $\Lambda$, there is the possibility of octet or flavor-singlet quantum numbers.
We consider three types 
of interpolating fields for the $\Lambda$: the octet $\Lambda$, 
\begin{eqnarray}
\eta^{ \Lambda_o}= 
2 \sqrt{1\over6}\, \epsilon^{abc}&& \left\{\left[
 2(u^{aT}C\gamma_5 d^b) s^c
+ (u^{aT}C\gamma_5 s^b) d^c
- (d^{aT}C\gamma_5 s^b) u^c
\right] \right. \nonumber \\& &  \left.
+ \beta \left[
 2(u^{aT}C d^b)\gamma_5 s^c
+ (u^{aT}C s^b)\gamma_5 d^c
- (d^{aT}C s^b)\gamma_5 u^c
\right] \right\},
\end{eqnarray}
the flavor-singlet $\Lambda$, 
\begin{eqnarray}
\eta^{ \Lambda_s}=
2 \sqrt{1\over3}\, \epsilon^{abc}&&\left\{ \left[
  (u^{aT}C\gamma_5 d^b) s^c
- (u^{aT}C\gamma_5 s^b) d^c
+ (d^{aT}C\gamma_5 s^b) u^c
\right] \right. \nonumber \\& &  \left.
+ \beta \left[
  (u^{aT}C d^b)\gamma_5 s^c
- (u^{aT}C s^b)\gamma_5 d^c
+ (d^{aT}C s^b)\gamma_5 u^c
\right]\right\},
\end{eqnarray}
and what we refer to as the common $\Lambda$,
\begin{equation}
\eta^{ \Lambda_c}=
2 \sqrt{1\over2}\, \epsilon^{abc} \left\{\left[
(u^{aT}C\gamma_5 s^b) d^c - (d^{aT}C\gamma_5 s^b) u^c
\right] + \beta \left[
(u^{aT}C s^b)\gamma_5 d^c - (d^{aT}C s^b)\gamma_5 u^c
\right] \right\}.
\end{equation}
which consists of terms common to $\eta^{ \Lambda_o}$ and $\eta^{ \Lambda_s}$.
Since SU(3) flavor is broken by the strange quark, it may be interesting to investigate
such an interpolating field which does not impose a flavor symmetry of the quarks
composing $\Lambda$.

The normalization in the above octet currents (excluding $\eta^{ \Lambda_s}$ and $\eta^{ \Lambda_c}$)
is chosen so that the correlation functions at the quark level coincide with 
each other under SU(3) flavor symmetry. This provides a simple check of the calculation.

For spin-3/2 non-strange N* with isospin 1/2, we consider:
\begin{equation}
\eta^{ N}_{3/2,\mu}=\epsilon^{abc}\left[
\left(u^{aT}C\sigma_{\rho\lambda} d^b\right)
\sigma^{\rho\lambda}\gamma_\mu u^c
- \left(u^{aT}C\sigma_{\rho\lambda} u^b\right)
\sigma^{\rho\lambda}\gamma_\mu d^c \right].
\label{nuc32}
\end{equation}
It has been known that the ground-state nucleon couples to 
this interpolating field via its spin 1/2 component.
As for the strange members, we consider for the $\Sigma$:
\begin{equation}
\eta^{ \Sigma}_{3/2,\mu}=\epsilon^{abc}\left[
\left(u^{aT}C\sigma_{\rho\lambda} s^b\right)
\sigma^{\rho\lambda}\gamma_\mu u^c
- \left(u^{aT}C\sigma_{\rho\lambda} u^b\right)
\sigma^{\rho\lambda}\gamma_\mu s^c \right],
\end{equation}
for the $\Xi$:
\begin{equation}
\eta^{ \Xi}_{3/2,\mu}=\epsilon^{abc}\left[
\left(s^{aT}C\sigma_{\rho\lambda} u^b\right)
\sigma^{\rho\lambda}\gamma_\mu s^c
- \left(s^{aT}C\sigma_{\rho\lambda} s^b\right)
\sigma^{\rho\lambda}\gamma_\mu u^c \right],
\end{equation}
and again for the three types of $\Lambda$:
\begin{eqnarray}
\eta^{ \Lambda_o}_{3/2,\mu}=\sqrt{1\over6}\, \epsilon^{abc}&&\left[
2\left(u^{aT}C\sigma_{\rho\lambda} d^b\right)
\sigma^{\rho\lambda}\gamma_\mu s^c
+ \left(u^{aT}C\sigma_{\rho\lambda} s^b\right)
\sigma^{\rho\lambda}\gamma_\mu d^c
\right.
\nonumber \\ & &
\left.
- \left(d^{aT}C\sigma_{\rho\lambda} s^b\right)
\sigma^{\rho\lambda}\gamma_\mu u^c \right],
\end{eqnarray}
\begin{eqnarray}
\eta^{ \Lambda_s}_{3/2,\mu}=\sqrt{1\over3}\, \epsilon^{abc}&&\left[
\left(u^{aT}C\sigma_{\rho\lambda} d^b\right)
\sigma^{\rho\lambda}\gamma_\mu s^c
- \left(u^{aT}C\sigma_{\rho\lambda} s^b\right)
\sigma^{\rho\lambda}\gamma_\mu d^c
\right.
\nonumber \\ & &
\left.
+ \left(d^{aT}C\sigma_{\rho\lambda} s^b\right)
\sigma^{\rho\lambda}\gamma_\mu u^c \right],
\end{eqnarray}
and
\begin{equation}
\eta^{ \Lambda_c}_{3/2,\mu}=\sqrt{1\over2}\, \epsilon^{abc}\left[
\left(u^{aT}C\sigma_{\rho\lambda} s^b\right)
\sigma^{\rho\lambda}\gamma_\mu d^c
- \left(d^{aT}C\sigma_{\rho\lambda} s^b\right)
\sigma^{\rho\lambda}\gamma_\mu u^c \right].
\end{equation}
This completes the list of all the interpolating fields considered in this work.

A baryon interpolating field excites both the ground state and the excited states 
with both parities.
The ability of an interpolating field to annihilate a baryon 
into the QCD vacuum is described by a phenomenological parameter 
(often referred to as current coupling or pole residue).
It is defined for the spin-1/2 interpolating field by
\begin{equation}
\langle 0\,|\,\eta_{1/2}\,|\,1/2^+ p s\rangle 
=\lambda_{1/2^+}\,u(p,s),
\end{equation}
for a positive-parity state with momentum $p$ and spin $s$, and 
\begin{equation}
\langle 0\,|\,\eta_{1/2}\,|\,1/2^- p s\rangle 
=\lambda_{1/2^-}\,\gamma_5\,u(p,s),
\end{equation}
for a negative-parity state.
Here $u$ is the Dirac spinor with the normalization 
$\bar{u}(p,s)\,u(p,s)=2M_B$, where $M_B$ is the baryon mass.

For the spin-3/2 interpolating field, it is defined by
\begin{equation}
\langle 0\,|\,\eta_{3/2, \mu}\,|\,3/2^+ ps\rangle =
\lambda_{3/2^+}\,u_\mu(p,s),
\end{equation}
for a positive-parity state, and 
\begin{equation}
\langle 0\,|\,\eta_{3/2, \mu}\,|\,3/2^- ps\rangle =
\lambda_{3/2^-}\,\gamma_5\, u_\mu(p,s),
\end{equation}
for a negative-parity state.
Here $u_\mu$ is a spin-vector in the Rarita-Schwinger representation
with the normalization $\bar{u_\mu}(p,s)\,u_\mu(p,s)=-2M_B$.
In addition, the coupling of a spin-3/2 current with 
a spin-1/2 state is defined by
\begin{equation}
\langle 0\,|\,\eta_{3/2, \mu}\,|\,1/2^+ ps\rangle =
\alpha_{1/2^+}\,\left({4p_\mu\over
M_B}+\gamma_\mu\right)\,\gamma_5\,u(p,s),
\end{equation}
for a positive-parity state, and 
\begin{equation}
\langle 0\,|\,\eta_{3/2, \mu}\,|\,1/2^- ps\rangle =
\alpha_{1/2^-}\,\left({4p_\mu\over M_B}+\gamma_\mu\right)\,u(p,s),
\end{equation}
for a negative-parity state.
Note that $\lambda$, $\alpha$ are two different parameters coupling to 
spin-1/2 states.
Knowledge of these parameters is helpful for 
three-point calculations in the QCD sum rule approach,
since it enters as normalization.

The QCD sum rules are derived using standard implementation of the 
approach. On the QCD side (conventionally referred to as the LHS), 
the correlator in~(\ref{cf2pt}) is evaluated 
using the Operator-Product-Expansion (OPE). 
On the phenomenological (or the RHS), the correlator is populated by intermediate states.
The Borel transform is applied to the momentum-space correlator to simultaneously 
provide exponential suppression of excited states and factorial suppression of 
higher-dimension operators. Furthermore a pole plus continuum model is usually adopted to 
reduce the number of phenomenological parameters to three: 
the ground-state pole mass ($M_B$), the current coupling ($\lambda^2$)
and the continuum threshold ($w$).
By numerically matching both sides of the sum rule over a reasonable window 
in the Borel mass parameter, these parameters are extracted as predictions of the method.
Whether a sum rule is saturated by a positive-parity state or 
a negative-parity state can be determined by examining the leading-order terms 
in the LHS and the sign of RHS.

The standard treatment of the QCD side proceeds by inserting the explicit
forms of the interpolating fields into the two-point function of Eq.~(\ref{cf2pt}) 
and contracting out pairs of time-ordered quark-field operators
which are the fully-interacting quark propagators of QCD.
The result is a so-called master formula expressed in terms of the quark propagators.
For completeness, all of the master formulas used in this work are given in the 
Appendix.

The fully-interacting quark propagator can be
derived from the Wick's theorem for the time-ordered product of two fermion fields
under the usual assumption of vacuum saturation of the intermediate
states of composite operators. The result, which includes operators up to dimension 8 
and terms linear in the quark mass, is explicitly given by
\begin{eqnarray}
S^{ab}_q(x,0) &\equiv& \lefteqn{\langle \Omega \bigm | T \left \{ q^a (x)
\overline q^b (0) \right \} \bigm | \Omega \rangle = } \hspace{36pt}
\nonumber \\
         && {i \over 2 \pi^2 x^4} \, \gamma \cdot x \, \delta^{ab}
             -{ m_q \over 2^2 \pi^2 x^2} \delta^{ab}
             -{1 \over 2^2 \, 3} \langle \overline q q \rangle
              \delta^{ab} \nonumber \\
         && +{i \over 2^4 \, 3} m_q \langle \overline q q \rangle
           \, \gamma \cdot x \, \delta^{ab}
          +{x^2 \over 2^6 \, 3} \langle \overline q g_c \sigma
           \cdot G q \rangle \delta^{ab} \nonumber \\
         && -{i x^2 \over 2^7 \, 3^2} m_q
\langle \overline q g_c \sigma \cdot G q \rangle \, \gamma \cdot x \,
\delta^{ab}
          -{ x^4 \over 2^{10} \, 3^3} \langle \overline q q \rangle
\langle g_c^2 G^2 \rangle \delta^{ab}  \nonumber \\
         && + {i \over 2^5 \pi^2 x^2}
\bigl ( g_c G_{\alpha \beta}^n \bigr )
\bigl (  \gamma \cdot x \, \sigma^{\alpha \beta} + \sigma^{\alpha \beta}
\, \gamma \cdot x  \bigr )
{\lambda^n_{ab} \over 2} \nonumber \\
&& + {1 \over 2^5 \pi^2} m_q \left [
\ln \left ( {-x^2 \Lambda^2 \over 4} \right ) + 2 \gamma_{EM} \right ]
\bigl ( g_c G_{\alpha \beta}^n \bigr )
\sigma^{\alpha \beta}
{\lambda^n_{ab} \over 2} \, 
\nonumber \\
&& - { 1 \over 2^6 \, 3} \langle \overline q g_c \sigma \cdot G q \rangle
\sigma_{\alpha \beta}
{\lambda^n_{ab} \over 2}  \nonumber \\
&& + {i \over 2^8 \, 3} m_q \langle \overline q g_c \sigma \cdot
G q \rangle
\bigl (  \gamma \cdot x \, \sigma_{\alpha \beta} + \sigma_{\alpha \beta}
\, \gamma \cdot x  \bigr )
{\lambda^n_{ab} \over 2} \nonumber \\
&& + {x^2 \over 2^{10} \, 3^2} \langle g_c^2 G^2 \rangle
\langle \overline q q \rangle
\sigma_{\alpha \beta}
{\lambda^n_{ab} \over 2} \, .
\label{qprop}
\end{eqnarray}

Since we deal with two independent interpolating fields for spin-1/2 and spin-3/2,
we consider three possible types of correlations:
the correlator of generalized spin-1/2 currents 
$\eta_{1/2}$ and $\bar{\eta}_{1/2}$,
the mixed correlator of generalized spin-1/2 current 
$\eta_{1/2, \mu} =\gamma_\mu\gamma_5\, \eta_{1/2}$
and spin-3/2 current $\bar{\eta}_{3/2, \nu}$,
and the correlator of spin-3/2 currents 
$\eta_{3/2, \mu}$ and $\bar{\eta}_{3/2, \nu}$.
From the 3 types of correlators, 11 independent sum rules can be constructed
which contain information on the lowing-lying states in 
the six channels ($N$, $\Sigma$, $\Xi$, $\Lambda_o$, $\Lambda_s$, $\Lambda_c$)
and in the spin-1/2$\pm$ and spin-3/2$\pm$ sectors.
The resulting sum rules are given in the following sections 
in sufficient detail so that they can be checked by other researchers.

\begin{figure}
\centerline{\psfig{file=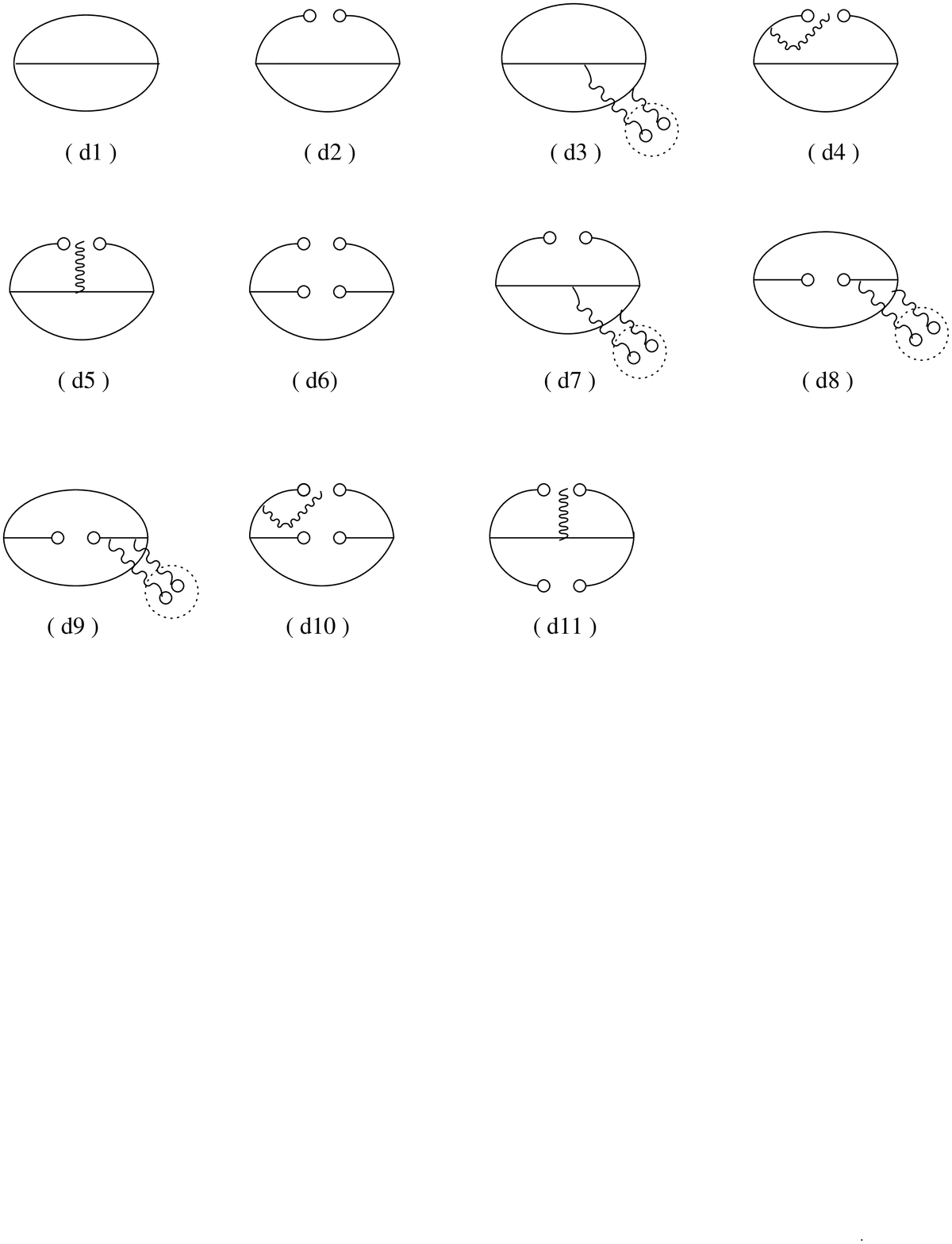,width=14.0cm}}
\vspace*{-8.5cm}
\caption{Skeleton diagrams considered in the calculation of the QCD side for a baryon.}
\label{diag0}
\end{figure}

\begin{figure}
\centerline{\psfig{file=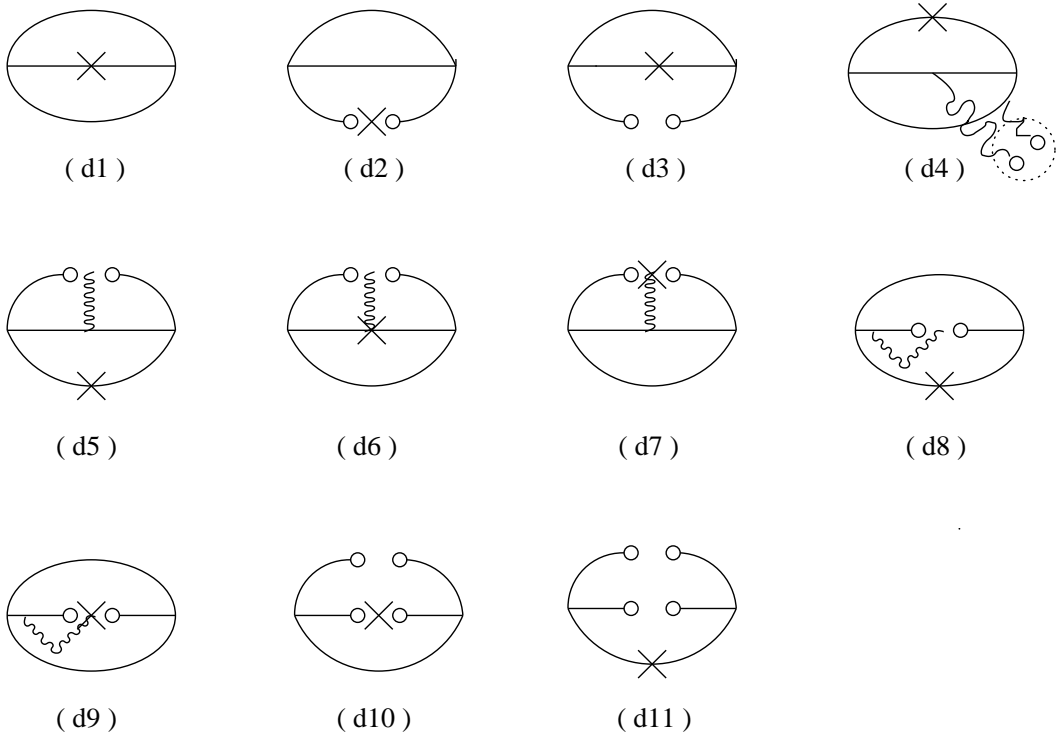,width=14.0cm,angle=-90}}
\vspace*{-8.5cm}
\caption{Skeleton diagrams linear in the quark mass, denoted by the cross symbol.}
\label{diag1}
\end{figure}

The quark propagator is then substituted into the master formula to give the LHS of the sum rule.
This tedious work can be organized by the 22 diagrams in Fig.~\ref{diag0} and Fig.~\ref{diag1} 
and is aided by the computer-algebra package REDUCE.
%

\subsection{QCD Sum Rules: Spin-1/2 Correlator}
\label{qcd11}

In this section we give the QCD sum rules as derived from 
the correlator of general spin-1/2 interpolating fields
$\eta_{1/2}$ and $\bar{\eta}_{1/2}$.

The phenomenological side of the correlation function can be written as
\begin{equation}
\Pi(p)= {\lambda^2_{1/2} \over p^2-M_B^2}\;(\hat{p}\pm M_B)+\cdots,
\end{equation}
where the dots denote excited state contributions.
The caret notation denotes $\hat{p}=p^\alpha\,\gamma_\alpha$.
The plus/minus sign corresponds to positive/negative parity.

The correlation function has two tensor structures, $\hat{p}$ and $1$,
from which two independent QCD sum rules can be constructed for each member of the octet family. 
It turns out that the QCD sum rules from the structure $\hat{p}$ contain
dimension-even condensates only, so we will refer such QCD sum rules as chiral-even.
Similarly, we refer to those from the structure $1$ as chiral-odd, 
because they involve dimension-odd condensates only. 
This classification scheme will be used throughout this work.

The chiral-even sum rules coupling to spin-1/2 states are
(the tensor structure is indicated on the left)
\begin{eqnarray}
& \hat{p}\, :\hspace{5mm} &
  c_1\; L^{-4/9}\; E_2\; M^6
+ c_2\; b\; L^{-4/9}\; E_0\; M^2
+ c_3\; m_s\,a\; L^{-4/9}\; E_0\; M^2
\nonumber \\ & &
+ c_4\; m_s\,m^2_0 a\; L^{-26/27}
+ c_5\; \kappa_v a^2\; L^{4/9}
+ c_6\; m^2_0 a^2\; L^{-2/27}\; {1\over M^2}
\nonumber \\ & &
= \tilde{\lambda}_{1/2^+}^2\; e^{-M^2_{1/2^+}/M^2} 
 +\tilde{\lambda}_{1/2^-}^2\; e^{-M^2_{1/2^-}/M^2},
\label{sum11a}
\end{eqnarray}
where the coefficients are given by
\begin{eqnarray}
& N:\hspace{5mm} &
c_1={1\over 16}(5+2\beta+5\beta^2),\;
\nonumber \\ & &
c_2= {1\over 64}(5+2\beta+5\beta^2),\;
\nonumber \\ & &
c_3= 0,\;
\nonumber \\ & &
c_4= 0,\;
\nonumber \\ & &
c_5= {1\over 6}[7-2\beta-5\beta^2],\;
\nonumber \\ & &
c_6= - {1\over 24}[13-2\beta-11\beta^2],
\label{nuc11a}
\end{eqnarray}
\begin{eqnarray}
& \Sigma:\hspace{5mm} &
c_1={1\over 16}(5+2\beta+5\beta^2),\;
\nonumber \\ & &
c_2= {1\over 64}(5+2\beta+5\beta^2),\;
\nonumber \\ & &
c_3= {1\over 8}[(12-5f_s)-2f_s\beta-(12+5f_s)\beta^2],\;
\nonumber \\ & &
c_4= - {1\over 24}[(4f_s+21+18t(M^2))+4f_s\beta
+(4f_s-21-18t(M^2))\beta^2],\;
\nonumber \\ & &
c_5= {1\over 6}[(6f_s+1)-2\beta-(6f_s-1)\beta^2],\;
\nonumber \\ & &
c_6= - {1\over 24}[(12f_s+1)-2\beta-(12f_s-1)\beta^2],
\label{sig11a}
\end{eqnarray}
\begin{eqnarray}
& \Xi:\hspace{5mm} &
c_1=  {1\over 16}(5+2\beta+5\beta^2),\;
\nonumber \\ & &
c_2= {1\over 64}(5+2\beta+5\beta^2),\;
\nonumber \\ & &
c_3= {3\over 4}[(2-f_s)-2f_s\beta-(2+f_s)\beta^2],\;
\nonumber \\ & &
c_4= - {1\over 24}[(15-f_s+18t(M^2))-10f_s\beta
-(15+f_s+18t(M^2))\beta^2],\;
\nonumber \\ & &
c_5= {f_s\over 6}[(f_s+6)-2f_s\beta+(f_s-6)\beta^2],\;
\nonumber \\ & &
c_6= - {f_s\over 24}[(f_s+12)-2f_s\beta+(f_s-12)\beta^2],
\label{xi11a}
\end{eqnarray}
\begin{eqnarray}\;
& \Lambda_o:\hspace{5mm} &
c_1={1\over 16}(5+2\beta+5\beta^2),\;
\nonumber \\ & &
c_2= {1\over 64}(5+2\beta+5\beta^2),\;
\nonumber \\ & &
c_3= {1\over 24}[(20-15f_s)-(16+6f_s)\beta-(4+15f_s)\beta^2],\;
\nonumber \\ & &
c_4= {1\over 24}[(4f_s-5-6t(M^2))+(4+4f_s)\beta
+(4f_s+1+6t(M^2))\beta^2],\;
\nonumber \\ & &
c_5= {1\over 18}[(10f_s+11)+(2-8f_s)\beta-(2f_s+13)\beta^2],\;
\nonumber \\ & &
c_6= {1\over 72}[(-16f_s-23)+(8f_s-2)\beta+(8f_s+25)\beta^2],
\label{lamo11a}
\end{eqnarray}
\begin{eqnarray}\;
& \Lambda_s:\hspace{5mm} &
c_1={1\over 8}(1-2\beta+\beta^2),\;
\nonumber \\ & &
c_2= {1\over 32}(1-2\beta+\beta^2),\;
\nonumber \\ & &
c_3= {1\over 12}[(-4-3f_s)+(8+6f_s)\beta-(4+3f_s)\beta^2],\;
\nonumber \\ & &
c_4= {1\over 12}(1-2\beta+\beta^2),\;
\nonumber \\ & &
c_5= {1\over 9}[(-1-2f_s)+(2+4f_s)\beta-(1+2f_s)\beta^2],\;
\nonumber \\ & &
c_6= {1\over 36}[(1+2f_s)-(4f_s+2)\beta+(1+2f_s)\beta^2],
\label{lams11a}
\end{eqnarray}
\begin{eqnarray}\;
& \Lambda_c:\hspace{5mm} &
c_1={1\over 16}(3-2\beta+3\beta^2),\;
\nonumber \\ & &
c_2= {1\over 64}(3-2\beta+3\beta^2),\;
\nonumber \\ & &
c_3= {1\over 8}[(4-3f_s)+2f_s\beta-(4+3f_s)\beta^2],\;
\nonumber \\ & &
c_4= {1\over 24}[(-3-6t(M^2))+4f_s\beta
+(3+6t(M^2))\beta^2],\;
\nonumber \\ & &
c_5= {1\over 6}[(2f_s-1)+2\beta-(2f_s+1)\beta^2],\;
\nonumber \\ & &
c_6= {1\over 24}[(-4f_s+1)-2\beta+(4f_s+1)\beta^2],
\label{lamc11a}
\end{eqnarray}

Herein after, the condensates and couplings are re-scaled as 
$a=-(2\pi)^2\,\langle\bar{u}u\rangle$,
$b=\langle g^2_c\, G^2\rangle$, $\langle\bar{u}g_c\sigma\cdot G
u\rangle=-m_0^2\,\langle\bar{u}u\rangle$,
$\tilde{\lambda}=(2\pi)^2\lambda$, 
$\tilde{\alpha}=(2\pi)^2\alpha$. 
The ratio $f_s=\langle\bar{s}s\rangle/\langle\bar{u}u\rangle
=\langle\bar{s}g_c\sigma\cdot G s\rangle
/\langle\bar{u}g_c\sigma\cdot G u\rangle$ accounts for 
the flavor symmetry breaking of the strange quark.
We retain only terms linear in the strange quark mass and set $m_u=m_d=0$.
The four-quark condensate is not well-known and we use the 
factorization approximation 
$\langle\bar{u}u\bar{u}u\rangle=\kappa_v\langle\bar{u}u\rangle^2$ and 
investigate its possible violation via the parameter $\kappa_v$.
The anomalous dimension corrections of the various operators 
are taken into account via the factors $L^\gamma=\left[{\alpha_s(\mu^2)/
\alpha_s(M^2)}\right]^\gamma =\left[{\ln(M^2/\Lambda_{QCD}^2)/
\ln(\mu^2/\Lambda_{QCD}^2)}\right]^\gamma$, where $\gamma$ is the 
appropriate anomalous dimension, $\mu=500$ MeV is the
renormalization scale, and $\Lambda_{QCD}$ is the QCD scale parameter 
whose value will be given below.
The function $t(M^2)\equiv\ln{M^2\over \mu^2}-\gamma_{EM}$
with $\gamma_{EM}\approx 0.577$ the Euler-Mascheroni constant. 
As usual, the excited state contributions are modeled using terms 
on the OPE side surviving $M^2\rightarrow \infty$ under the assumption
of duality, and are represented by the factors 
$E_n(x)=1-e^{-x}\sum_n{x^n/n!}$ on the LHS with $x=w_B^2/M_B^2$
and $w_B$ an effective continuum threshold. Note that $w_B$
is in general different from sum rule to sum rule and we will 
treat it as a free parameter in the Monte-Carlo analysis.

A trivial check of the calculation is provided by the fact that the QCD sum rules 
for $\Sigma$ should reduce to those for the nucleon by replacing the s-quark with 
the d-quark, which is equivalent to setting $m_s=0$ and $f_s=1$.
This check is performed throughout the work.

The chiral-odd sum rules coupling to spin-1/2 states at the structure $1$ are
\begin{eqnarray}
& 1\,:\hspace{5mm} &
c_1\; m_s\; L^{-8/9}\; E_2\; M^6
+c_2\; a\; E_1\; M^4
+c_3\; m^2_0 a\; L^{-14/27}\; E_0\; M^2
\nonumber \\ & &
+c_4\; m_s \, b \; L^{-8/9}\; E_0\; M^2
+c_5\; a\, b
+c_6\; m_s\, \kappa_v a^2
\nonumber \\ & &
=\tilde{\lambda}_{1/2^+}^2\; M_{1/2^+}\; e^{-M^2_{1/2^+}/M^2} 
 - \tilde{\lambda}_{1/2^-}^2\; M_{1/2^-}\; e^{-M^2_{1/2^-}/M^2},
\label{sum11b}
\end{eqnarray}
where the coefficients are given by
\begin{eqnarray}
& N:\hspace{5mm} &
c_1= 0,\;
c_2= {1\over 4}[7-2\beta-5\beta^2)],\;
c_3= - {3\over 4}(1-\beta^2),\;
c_4= 0,\;
\nonumber \\ & &
c_5=  {1\over 288}[19+10\beta-29\beta^2],\;
c_6= 0,
\label{nuc11b}
\end{eqnarray}
\begin{eqnarray}
& \Sigma:\hspace{5mm} &
c_1=  {1\over 4}(1-\beta)^2,\;
c_2=  {1\over 4}[(6+f_s)-2f_s\beta-(6-f_s)\beta^2)],\;
c_3= - {3\over 4}(1-\beta^2),\;
\nonumber \\ & &
c_4= - {1\over 32}(1-\beta)^2,\;
c_5=  {1\over 288}[(24-5f_s)+10f_s\beta-(24+5f_s)\beta^2],\;
\nonumber \\ & &
c_6=  {1\over 6}[(5-3f_s)+2\beta+(5+3f_s)\beta^2],
\label{sig11b}
\end{eqnarray}
\begin{eqnarray}
& \Xi:\hspace{5mm} &
c_1=  {3\over 2}(1-\beta^2),\;
c_2=  {1\over 4}[(6f_s+1)-2\beta-(6f_s-1)\beta^2)],\;
c_3= - {3f_s\over 4}(1-\beta^2),\;
\nonumber \\ & &
c_4=  {3\over 16}(1-\beta^2),\;
c_5=  {1\over 288}[(24f_s-5)+10\beta-(24f_s+5)\beta^2],\;
\nonumber \\ & &
c_6=  {f_s\over 2}[(3-f_s)+2\beta+(3+f_s)\beta^2],
\label{xi11b}
\end{eqnarray}
\begin{eqnarray}
& \Lambda_o:\hspace{5mm} &
c_1=  {1\over 12}(11+2\beta-13\beta^2),\;
c_2=  {1\over 12}[(10+11f_s)+(-8+2f_s)\beta-(2+13f_s)\beta^2)],\;
\nonumber \\ & &
c_3= {1\over 4}((-1-2f_s)+(1+2f_s)\beta^2),\;
c_4= {1\over 96}(13-2\beta-11\beta^2),\;
\nonumber \\ & &
c_5=  {1\over 864}[(4+53f_s)+(40-10f_s)\beta-(44+43f_s)\beta^2],\;
\nonumber \\ & &
c_6=  {1\over 18}[(15-5f_s)+(6+4f_s)\beta+(15+f_s)\beta^2],
\label{lamo11b}
\end{eqnarray}
\begin{eqnarray}
& \Lambda_s:\hspace{5mm} &
c_1=  {1\over 6}(-1+2\beta-\beta^2),\;
c_2=  {1\over 6}[(-2-f_s)+(4+2f_s)\beta-(2+f_s)\beta^2)],\;
\nonumber \\ & &
c_3= 0\;
c_4= {1\over 48}(1-2\beta+\beta^2),\;
\nonumber \\ & &
c_5= {1\over 432}[(10+5f_s)-(20+10f_s)\beta+(10+5f_s)\beta^2],\;
\nonumber \\ & &
c_6=  {1\over 9}[(3+f_s)-(6+2f_s)\beta+(3+f_s)\beta^2],
\label{lams11b}
\end{eqnarray}
\begin{eqnarray}
& \Lambda_c:\hspace{5mm} &
c_1=  {1\over 4}(-1+2\beta-\beta^2),\;
c_2=  {1\over 4}[(2-f_s)+2f_s\beta-(2+f_s)\beta^2)],\;
\nonumber \\ & &
c_3= {1\over 4}(-1+\beta^2),\;
c_4= {1\over 32}(1-2\beta+\beta^2),\;
\nonumber \\ & &
c_5=  {1\over 288}[(8+5f_s)-10f_s\beta+(-8+5f_s)\beta^2],\;
\nonumber \\ & &
c_6=  {1\over 6}[(3-f_s)-2\beta+(3+f_s)\beta^2],
\label{lamc11b}
\end{eqnarray}
Note that on the phenomenological side of the sum rules here and below, 
we explicitly retain the two lowest states,
one having positive-parity and one having negative-parity, 
because it is not known {\it apriori} whether a sum rule is 
saturated by a positive or negative state.
Only one term is used in the analysis once it is determined that 
it is the dominant one.

\subsection{QCD Sum Rules: Mixed Correlator}
\label{qcd13}

In this section we give the QCD sum rules as derived from 
the mixed correlator of generalized spin-1/2 current 
$\eta_{1/2, \mu} =\gamma_\mu\gamma_5\, \eta_{1/2}$
and spin-3/2 current $\bar{\eta}_{3/2, \nu}$.

The phenomenological side of the correlation function can be written as
\begin{equation}
\Pi_{ \mu\nu}(p)=
{-\lambda_{1/2}\alpha_{1/2} \over p^2-M_B^2}
\left\{\gamma_\mu\gamma_\nu\hat{p} \pm M_B\gamma_\mu\gamma_\nu
\mp {4\over M_B} \gamma_\mu p_\nu \hat{p}
+ 2 \gamma_\mu p_\nu \right\} + \cdots,
\end{equation}
where the dots denote excited state contributions.
The upper/lower sign corresponds to positive/negative parity.
There are four tensor structures from which four QCD sum rules can be 
constructed for each member. 
It turns out that the sum rules from the structures 
$\gamma_\mu p_\nu \hat{p}$ and $\gamma_\mu p_\nu$ are identical,
leading to three independent sum rule sets.

The chiral-even sum rules coupling to spin-1/2 states are
\begin{eqnarray}
& \gamma_\mu\gamma_\nu\hat{p}\, : \hspace{5mm} &
c_1\; m_s\,a\; L^{-4/27}\; E_0\; M^2
+c_2\; \kappa_v a^2\; L^{20/27}
+c_3\; m_s\,m^2_0 a\; L^{-18/27}
+c_4\; m^2_0 a^2\; L^{2/9}\; {1\over M^2}
\nonumber \\ & &
= \,-\tilde{\lambda}_{1/2^+}\;\tilde{\alpha}_{1/2^+}\;
e^{-M^2_{1/2^+}/M^2}
 -\,\tilde{\lambda}_{1/2^-}\;\tilde{\alpha}_{1/2^-}\;
e^{-M^2_{1/2^-}/M^2},
\label{sum13a}
\end{eqnarray}
where the coefficients are given by
\begin{eqnarray}
& N:\hspace{5mm} &
c_1= 0,\;
c_2= -2,\;
c_3= 0,\; 
c_4= {1\over 36}[27-3\beta],
\label{nuc13a}
\end{eqnarray}
\begin{eqnarray}
& \Sigma:\hspace{5mm} &
c_1= -{3-\beta\over 2},\;
c_2= -{1\over 3}[3(1+f_s)+(1-f_s)\beta],\;
\nonumber \\ & &
c_3= {1\over 24}[(21+2f_s)-2(1+\beta)\,t(M^2)-(7+2f_s)\beta],\; 
\nonumber \\ & &
c_4= {1\over 36}[(13+14f_s)-3(2f_s-1)\beta],
\label{sig13a}
\end{eqnarray}
\begin{eqnarray}
& \Xi:\hspace{5mm} &
c_1= -{1\over 2}[3(2f_s+1)+(2f_s-1)\beta],\;
c_2= -{f_s\over 3}[3(1+f_s)+(f_s-1)\beta],\;
\nonumber \\ & &
c_3= {1\over 24}[(7+24f_s)-2(1+\beta)(1+4f_s)\,t(M^2)-(5-8f_s)\beta],\; 
\nonumber \\ & &
c_4= {f_s\over 36}[(14+13f_s)-3(2-f_s)\beta],
\label{xi13a}
\end{eqnarray}
\begin{eqnarray}
& \Lambda_o:\hspace{5mm} &
c_1= -{5+\beta\over 6},\;
c_2= -{1\over 9}[(1+5f_s)-(1-f_s)\beta],\;
\nonumber \\ & &
c_3= {1\over 72}[(17-6t(M^2))+(5-6t(M^2))\beta],\; 
\nonumber \\ & &
c_4= {1\over 108}[(5+22f_s)+(-5+2f_s)\beta],
\label{lamo13a}
\end{eqnarray}
\begin{eqnarray}
& \Lambda_s:\hspace{5mm} &
c_1= -{-1+\beta\over 3},\;
c_2= -{1\over 9}[2(1-f_s)+2(f_s-1)\beta],\;
\nonumber \\ & &
c_3= {1\over 36}(-1+5\beta),\; 
\nonumber \\ & &
c_4= {1\over 54}[5(1-f_s)+5(-1+f_s)\beta],
\label{lams13a}
\end{eqnarray}
\begin{eqnarray}
& \Lambda_c:\hspace{5mm} &
c_1= -{1+\beta\over 2},\;
c_2= -{1\over 3}[(1+f_s)+(-1+f_s)\beta],\;
\nonumber \\ & &
c_3= {1\over 24}[(5-2t(M^2))+(5-2t(M^2))\beta],\; 
\nonumber \\ & &
c_4= -{1\over 36}[(5+4f_s)+(-5+4f_s)\beta],
\label{lamc13a}
\end{eqnarray}
An interesting feature of this chiral-even sum rule 
is that there is only a relatively small term of the order $m_s M^2$ on the OPE
to model the continuum for the strange baryons, while no such term for the nucleon.
One consequence is that the continuum contribution is $zero$ for the nucleon and
small for the strange baryons, suggesting a weak coupling between excited states and OPE.
It makes it difficult to select a valid Borel region independently to fit the sum rule. 

The chiral-odd sum rules coupling to spin-1/2 states are
\begin{eqnarray}
& \gamma_\mu\gamma_\nu\; : \hspace{5mm} &
c_1\; m_s\; L^{-16/27}\; E_2\; M^6
+ c_2\;  a\; L^{8/27}\; E_1\; M^4
+ c_3\; a\,b\; L^{8/27}
+ c_4\; m_s \kappa_v a^2\; L^{8/27}
\nonumber \\ & &
= - \tilde{\lambda}_{1/2^+}\;\tilde{\alpha}_{1/2^+}\; M_{1/2+}\; 
     e^{-M^2_{1/2^+}/M^2}
 + \tilde{\lambda}_{1/2^-}\;\tilde{\alpha}_{1/2^-}\; M_{1/2-}\; 
     e^{-M^2_{1/2^-}/M^2},
\label{sum13b}
\end{eqnarray}
where the coefficients are given by
\begin{eqnarray}
& N:\hspace{5mm} &
c_1= 0,\;
c_2= -1,\;
c_3= -{1\over 144}[3+9\beta],\; 
c_4= 0,
\label{nuc13b}
\end{eqnarray}
\begin{eqnarray}
& \Sigma:\hspace{5mm} &
c_1= -{3+\beta\over 4},\;
c_2= -{1\over 6}[3(1+f_s)-(1-f_s)\beta],\;
\nonumber \\ & &
c_3= -{1\over 144}[3f_s+(4+5f_s)\beta],\; 
c_4= -{f_s\over 6}(3-\beta),
\label{sig13b}
\end{eqnarray}
\begin{eqnarray}
& \Xi:\hspace{5mm} &
c_1= -{3-\beta\over 4},\;
c_2= -{1\over 6}[3(1+f_s)+(1-f_s)\beta],\;
\nonumber \\ & &
c_3= -{1\over 144}[3+(5+4f_s)\beta],\; 
c_4= -{f_s\over 6}[3(2+f_s)+(2-f_s)\beta],
\label{xi13b}
\end{eqnarray}
\begin{eqnarray}
& \Lambda_o:\hspace{5mm} &
c_1= -{1-\beta\over 12},\;
c_2= -{1\over 18}[(5+f_s)+(1-f_s)\beta],\;
\nonumber \\ & &
c_3= -{1\over 432}[(4-f_s)+(8+f_s)\beta],\; 
c_4= -{f_s\over 18}(5+\beta),
\label{lamo13b}
\end{eqnarray}
\begin{eqnarray}
& \Lambda_s:\hspace{5mm} &
c_1= -{1-\beta\over 6},\;
c_2= -{1\over 9}[(-1+f_s)+(1-f_s)\beta],\;
\nonumber \\ & &
c_3= -{1\over 216}[(1-f_s)+(-1+f_s)\beta],\; 
c_4= -{f_s\over 9}(-1+\beta),
\label{lams13b}
\end{eqnarray}
\begin{eqnarray}
& \Lambda_c:\hspace{5mm} &
c_1= -{1-\beta\over 4},\;
c_2= -{1\over 6}[(1+f_s)+(1-f_s)\beta],\;
\nonumber \\ & &
c_3= -{1\over 144}[(2-f_s)+(2+f_s)\beta],\; 
c_4= -{f_s\over 6}(1+\beta),
\label{lamc13b}
\end{eqnarray}

The chiral-odd sum rules coupling to spin-1/2 states are 
\begin{eqnarray}
& \gamma_\mu p_\nu \hat{p}\; : \hspace{5mm} &
c_1\; m_s\; L^{-16/27}\; E_1\; M^4
+ c_2\;  a\; L^{8/27}\; E_0\; M^2
+ c_3\; m^2_0 a\; L^{-2/9}
\nonumber \\ & &
+ c_4\; m_s\,b\; L^{-16/27}
+ c_5\; a\,b\; L^{8/27}\; {1\over M^2}
+ c_6\; m_s \kappa_v a^2\; L^{8/27}\; {1\over M^2}
\nonumber \\ & &
= {\tilde{\lambda}_{1/2^+}\; \tilde{\alpha}_{1/2^+} \over
M_{1/2^+}}\;
    e^{-M^2_{1/2^+}/M^2}
 - {\tilde{\lambda}_{1/2^-}\; \tilde{\alpha}_{1/2^-} \over
M_{1/2^-}}\;
    e^{-M^2_{1/2^-}/M^2},
\label{sum13c}
\end{eqnarray}
where the coefficients are given by
\begin{eqnarray}
& N:\hspace{5mm} &
c_1= 0,\;
c_2= 1,\;
c_3= -{1\over 8}(3-\beta),\;
c_4= 0,\;
c_5= -{1\over 48}[1+3\beta],\; 
c_6= 0,
\label{nuc13c}
\end{eqnarray}
\begin{eqnarray}
& \Sigma:\hspace{5mm} &
c_1= {3+\beta\over 8},\;
c_2= {1\over 6}[3(1+f_s)-(1-f_s)\beta],\;
c_3= -{1\over 24}(5+4f_s-3\beta),\;
\nonumber \\ & &
c_4= -{1\over 96}(1+3\beta),\;
c_5= -{1\over 144}[3f_s+(4+5f_s)\beta],\; 
c_6= -{f_s\over 6}(3-\beta),
\label{sig13c}
\end{eqnarray}
\begin{eqnarray}
& \Xi:\hspace{5mm} &
c_1= {3-\beta\over 8},\;
c_2= {1\over 6}[3(1+f_s)+(1-f_s)\beta],\;
c_3= -{1\over 24}(4+5f_s-3f_s\beta),\;
\nonumber \\ & &
c_4= {1\over 96}(1-3\beta),\;
c_5= -{1\over 144}[3+(5+4f_s)\beta],\; 
c_6= -{f_s\over 6}[3(2+f_s)+(2-f_s)\beta],
\label{xi13c}
\end{eqnarray}
\begin{eqnarray}
& \Lambda_o:\hspace{5mm} &
c_1= {1-\beta\over 24},\;
c_2= {1\over 18}[(5+f_s)+(1-f_s)\beta],\;
c_3= -{1\over 72}[(7+2f_s)-(1+2f_s)\beta],\;
\nonumber \\ & &
c_4= {1\over 288}(1-\beta),\;
c_5= -{1\over 432}[(4-f_s)+(8+f_s)\beta],\; 
c_6= -{f_s\over 18}(5+\beta),
\label{lamo13c}
\end{eqnarray}
\begin{eqnarray}
& \Lambda_s:\hspace{5mm} &
c_1= {1-\beta\over 12},\;
c_2= {1\over 9}[(-1+f_s)+(1-f_s)\beta],\;
c_3= -{1\over 18}[(-1+f_s)+(1-f_s)\beta],\;
\nonumber \\ & &
c_4= {1\over 144}(1-\beta),\;
c_5= -{1\over 216}[(1-f_s)+(-1+f_s)\beta],\; 
c_6= -{f_s\over 9}(-1+\beta),
\label{lams13c}
\end{eqnarray}
\begin{eqnarray}
& \Lambda_c:\hspace{5mm} &
c_1= {1-\beta\over 8},\;
c_2= {1\over 6}[(1+f_s)+(1-f_s)\beta],\;
c_3= -{1\over 24}[(1+2f_s)+(1-2f_s)\beta],\;
\nonumber \\ & &
c_4= {1\over 96}(1-\beta),\;
c_5= -{1\over 144}[(2-f_s)+(2+f_s)\beta],\; 
c_6= -{f_s\over 6}(1+\beta),
\label{lamc13c}
\end{eqnarray}
Comparing with existing calculations for the mixed correlator,
the coefficients in the nucleon channel in the above three sum rules 
agree with those in Ref.~\cite{Derek96}.
The coefficients in the strange baryon channels are new.
This turns out to be a very favorable sum rule among the 11 sum rules in predicting the 
masses of spin-1/2 and positive-parity baryons.

\subsection{QCD Sum Rules: Spin-3/2 Correlator}
\label{qcd33}

In this section we give the QCD sum rules as derived from 
the correlator of spin-3/2 currents 
$\eta_{3/2, \mu}$ and $\bar{\eta}_{3/2, \nu}$.

The phenomenological side of the correlation function can be written as
\begin{eqnarray}
\Pi_{\mu\nu}(p) & = & \lambda^2_{3/2}\; \left[ -g_{\mu\nu}\hat{p}
+{1\over 3}(\gamma_\mu p_\nu+\gamma_\nu p_\mu)
+{2\over 3}{p_\mu p_\nu\hat{p} \over M^2_{3/2}} \mp M_{3/2} g_{\mu\nu} 
\pm {1\over 3}M_{3/2}\gamma_\mu\gamma_\nu
\pm {2\over 3}{p_\mu p_\nu \over M_{3/2}}
\right] \nonumber \\  
&+& \alpha^2_{1/2}\; \left[ -4(\gamma_\mu p_\nu+\gamma_\nu p_\mu)
+ {16\over M^2_{1/2}} p_\mu p_\nu\hat{p}
\mp M_{1/2}\; \gamma_\mu\gamma_\nu \mp {8 \over M_{1/2}} p_\mu p_\nu
\,\right]
+ \cdots,
\end{eqnarray}
where the dots denote excited state contributions.
The upper/lower sign corresponds to positive/negative parity.
The tensor structures are chosen because they are orthogonal 
to each other.
Six independent sum rules can be constructed from the correlator for each member: 
two for spin-3/2 states only, two for spin-1/2 states only, and two for both.

The chiral-even sum rules coupling to spin-3/2 states are
\begin{eqnarray}
& g_{\mu \nu} \hat{p}\; : \hspace{5mm} &
  c_1\; L^{4/27}\; E_2\; M^6
+ c_2\; b\; L^{4/27}\; E_0\; M^2
+ c_3\; m_s\,a\; L^{4/27}\; E_0\; M^2
\nonumber \\ & &
+ c_4\; m_s\,m^2_0 a\; L^{-10/27}
+ c_5\; \kappa_v a^2\; L^{28/27}
+ c_6\; m^2_0 a^2\; L^{14/27}\; {1\over M^2}
\nonumber \\ & &
= - \tilde{\lambda}_{3/2^-}^2\; e^{-M^2_{3/2^-}/M^2}
- \tilde{\lambda}_{3/2^+}^2\; e^{-M^2_{3/2^+}/M^2},
\label{sum33a}
\end{eqnarray}
where the coefficients are given by
\begin{eqnarray}
& N:\hspace{5mm} &
c_1=-{24\over 5},\;
c_2= {5\over 3},\;
c_3= 0,\;
c_4= 0,\;
c_5= 16,\;
c_6= -{28\over 3}, \;
\label{nuc33a}
\end{eqnarray}
\begin{eqnarray}
& \Sigma:\hspace{5mm} &
c_1=-{24\over 5},\;
c_2= {5\over 3},\;
c_3= 8(f_s+4),\;
c_4= -{2\over 3}(11f_s+28),\;
c_5= {16\over 3}(4f_s-1),\;
\nonumber \\ & &
c_6= -{28\over 9}(4f_s-1),\;
\label{sig33a}
\end{eqnarray}
\begin{eqnarray}
& \Xi:\hspace{5mm} &
c_1= -{24\over 5},\;
c_2= {5\over 3},\;
c_3= 32,\;
c_4= -{2\over 3}(5f_s+28),\;
c_5= {16\over 3}f_s(4-f_s),\;
\nonumber \\ & &
c_6= -{28\over 9}f_s(4-f_s),\;
\label{xi33a}
\end{eqnarray}
\begin{eqnarray}
& \Lambda_o:\hspace{5mm} &
c_1= -{8\over 5},\;
c_2= {5\over 9},\;
c_3= {8\over 3}f_s,\;
c_4= -{22\over 9}f_s,\;
c_5= {16\over 9},\;
c_6= -{28\over 27},\;
\label{lamo33a}
\end{eqnarray}
\begin{eqnarray}
& \Lambda_s:\hspace{5mm} &
c_1= -{8\over 5},\;
c_2= {5\over 9},\;
c_3= {8\over 3}f_s,\;
c_4= -{20\over 9}f_s,\;
c_5= {32\over 9},\;
c_6= -{56\over 27},\;
\label{lams33a}
\end{eqnarray}
\begin{eqnarray}
& \Lambda_c:\hspace{5mm} &
c_1= -{8\over 5},\;
c_2= {5\over 9},\;
c_3= {8\over 3}f_s,\;
c_4= -2f_s,\;
c_5= {16\over 3},\;
c_6= -{28\over 9},\;
\label{lamc33a}
\end{eqnarray}
In this sum rule, states with both positive- and negative-parity contribute with 
the same sign, which means that the excited states contamination is large. 
Therefore it makes it difficult to resolve which is the dominant contribution 
under the single pole-plus-continuum model. 
One could retain both terms (a two-pole plus continuum model), but it increases 
the total number of phenomenological parameters to five, which is too many for 
a sum rule to resolve.
As a result, no predictions can be made from this sum rule.

The chiral-odd sum rules coupling to spin-3/2 states are
\begin{eqnarray}
& g_{\mu \nu}\; : \hspace{5mm} &
c_1\; m_s\; L^{-8/27}\; E_2\; M^6
+c_2\; a\; E_1\; L^{16/27}\; M^4
+c_3\; m^2_0 a\; L^{2/27}\; E_0\; M^2
\nonumber \\ & &
+c_4\; m_s \,b\; L^{-8/27}\; E_0\; M^2
+c_5\; a\,b\; L^{16/27}
+c_6\; m_s\,\kappa_v a^2\; L^{16/27}
\nonumber \\ & &
=\tilde{\lambda}_{3/2^-}^2\; M_{3/2^-}\; e^{-M^2_{3/2^-}/M^2}
- \tilde{\lambda}_{3/2^+}^2\; M_{3/2^+}\; e^{-M^2_{3/2^+}/M^2}
,
\label{sum33b}
\end{eqnarray}
where the coefficients are given by
\begin{eqnarray}
& N:\hspace{5mm} &
c_1= 0,\;
c_2= 16,\;
c_3= -8,\;
c_4= 0,\;
c_5= {-2\over 3},\;
c_6= 0,\;
\label{nuc33b}
\end{eqnarray}
\begin{eqnarray}
& \Sigma:\hspace{5mm} &
c_1= -6,\;
c_2= {16\over 3}(4-f_s),\;
c_3= {-8\over 3}(4-f_s),\;
c_4= {1\over 2},\;
c_5= {-2\over 9}(4-f_s),\;
\nonumber \\ & &
c_6= -16,\;
\label{sig33b}
\end{eqnarray}
\begin{eqnarray}
& \Xi:\hspace{5mm} &
c_1= 24,\;
c_2= {16\over 3}(4f_s-1),\;
c_3= {-8\over 3}(4f_s-1),\;
c_4= -2,\;
c_5= {-2\over 9}(4f_s-1),\;
\nonumber \\ & &
c_6= -32f_s,\;
\label{xi33b}
\end{eqnarray}
\begin{eqnarray}
& \Lambda_o:\hspace{5mm} &
c_1= 2,\;
c_2= {16\over 9}f_s,\;
c_3= -{8\over 9}f_s,\;
c_4= -{1\over 6},\;
c_5= -{2\over 27}f_s,\;
c_6= -{16\over 3},\;
\label{lamo33b}
\end{eqnarray}
\begin{eqnarray}
& \Lambda_s:\hspace{5mm} &
c_1= 4,\;
c_2= {32\over 9}f_s,\;
c_3= -{16\over 9}f_s,\;
c_4= -{1\over 3},\;
c_5= -{4\over 27}f_s,\;
c_6= -{16\over 3},\;
\label{lams33b}
\end{eqnarray}
\begin{eqnarray}
& \Lambda_c:\hspace{5mm} &
c_1= 6,\;
c_2= {16\over 3}f_s,\;
c_3= -{8\over 3}f_s,\;
c_4= -{1\over 2},\;
c_5= -{2\over 9}f_s,\;
c_6= -{16\over 3},\;
\label{lamc33b}
\end{eqnarray}
In this sum rule, the leading terms ($c_1$,$c_2$) on the LHS are positive, 
which leads us to conclude that the sum rule is dominated by negative-parity states 
because they come in with a positive sign on the RHS while the positive-parity states 
have the opposite sign. 
The parity cancellation in the excited spectrum may lead to a clean extraction of 
$J^P={3\over 2}^-$ states from this sum rule.

The chiral-even sum rules coupling to spin-1/2 states are 
\begin{eqnarray}
& \gamma_\mu p_\nu+\gamma_\nu p_\mu\; : \hspace{5mm} &
  c_1\; L^{4/27}\; E_2\; M^6
+ c_2\; b\; L^{4/27}\; E_0\; M^2
+ c_3\; m_s\,a\; L^{4/27}\; E_0\; M^2
\nonumber \\ & &
+ c_4\; m_s\,m^2_0 a\; L^{-10/27}
+ c_5\; \kappa_v a^2\; L^{28/27}
+ c_6\; m^2_0 a^2\; L^{14/27}\; {1\over M^2}
\nonumber \\ & &
= - \tilde{\alpha}_{1/2^+}^2\; e^{-M^2_{1/2^+}/M^2}
 - \tilde{\alpha}_{1/2^-}^2\; e^{-M^2_{1/2^-}/M^2},
\label{sum33c}
\end{eqnarray}
where the coefficients are given by
\begin{eqnarray}
& N:\hspace{5mm} &
c_1= {3\over 10},\;
c_2= -{5\over 24},\;
c_3= 0,\;
c_4= 0,\;
c_5= -2,\;
c_6= {7 \over 6},\;
\label{nuc33c}
\end{eqnarray}
\begin{eqnarray}
& \Sigma:\hspace{5mm} &
c_1= {3\over 10},\;
c_2= -{5\over 24},\;
c_3= -(f_s+4),\;
c_4= {(3f_s+7) \over 3} ,\;
c_5= -{2(4f_s-1) \over 3} ,\nonumber \\ & &
c_6= {7(4f_s-1) \over 18},\;
\label{sig33c}
\end{eqnarray}
\begin{eqnarray}
& \Xi:\hspace{5mm} &
c_1= {3\over 10},\;
c_2= -{5\over 24},\;
c_3= -4,\;
c_4= {(5f_s+14)\over 6},\;
c_5= {{2f_s(f_s-4)}\over 3},
\nonumber \\ & &
c_6= -{7f_s(f_s-4)\over 18},\;
\label{xi33c}
\end{eqnarray}
\begin{eqnarray}
& \Lambda_o:\hspace{5mm} &
c_1= {1\over 10},\;
c_2= -{5\over 72},\;
c_3= -{1\over 3}f_s,\;
c_4= {1\over 3}f_s,\;
c_5= -{2\over 9},
c_6= {7\over 54},\;
\label{lamo33c}
\end{eqnarray}
\begin{eqnarray}
& \Lambda_s:\hspace{5mm} &
c_1= {1\over 10},\;
c_2= -{5\over 72},\;
c_3= -{1\over 3}f_s,\;
c_4= {1\over 3}f_s,\;
c_5= -{4\over 9},
c_6= {7\over 27},\;
\label{lams33c}
\end{eqnarray}
\begin{eqnarray}
& \Lambda_c:\hspace{5mm} &
c_1= {1\over 10},\;
c_2= -{5\over 72},\;
c_3= -{1\over 3}f_s,\;
c_4= {1\over 3}f_s,\;
c_5= -{2\over 3},
c_6= {7\over 18},\;
\label{lamc33c}
\end{eqnarray}
Note that the RHS is always negative, but the leading term ($M^6$) on the LHS side
is positive. This is an indication of a poor sum rule.
In fact, no physical results have been extracted from this sum rule.

The chiral-odd sum rules coupling to spin-1/2 states are 
\begin{eqnarray}
& \gamma_\mu\gamma_\nu + {1\over 3}\, g_{\mu \nu}\; : \hspace{5mm} &
c_1\; a\; L^{16/27}\; E_1\; M^4
+c_2\; m^2_0 a\; L^{2/27}\; E_0\; M^2
\nonumber \\ & &
+c_3\; m_s \,b\; L^{-8/27}\; E_0\; M^2
+c_4\; a\,b\; L^{16/27}
+c_5\; m_s\,\kappa_v a^2\; L^{16/27}
\nonumber \\ & &
= - \tilde{\alpha}_{1/2^+}^2\; M_{1/2+}\; e^{-M^2_{1/2^+}/M^2}
 + \tilde{\alpha}_{1/2^-}^2\; M_{1/2-}\; e^{-M^2_{1/2^-}/M^2},
\label{sum33d}
\end{eqnarray}
where the coefficients are given by
\begin{eqnarray}
& N:\hspace{5mm} &
c_1= {1\over 3},\;
c_2= -{2\over 3},\;
c_3= 0,\;
c_4= {5\over 72},\;
c_5= 0,\;
\label{nuc33d}
\end{eqnarray}
\begin{eqnarray}
& \Sigma:\hspace{5mm} &
c_1= -{1\over 9}(f_s-4),\;
c_2= {2\over 9}(f_s-4),\;
c_3= {1\over 24},\;
c_4= -{5\over 216}(f_s-4),\;
\nonumber \\ & &
c_5= -{4\over 3}(1+f_s),\;
\label{sig33d}
\end{eqnarray}
\begin{eqnarray}
& \Xi:\hspace{5mm} &
c_1= {1\over 9}(4 f_s-1),\;
c_2= -{2\over 9}(4 f_s-1),\;
c_3= -{1\over 6},\;
c_4= {5\over 216}(4 f_s-1),\;
\nonumber \\ & &
c_5= -{2\over 3}f_s(3+2f_s),\;
\label{xi33d}
\end{eqnarray}
\begin{eqnarray}
& \Lambda_o:\hspace{5mm} &
c_1= {1\over 27} f_s,\;
c_2= -{2\over 27} f_s,\;
c_3= -{1\over 72},\;
c_4= {5\over 648} f_s,\;
c_5= -{4\over 9},\;
\label{lamo33d}
\end{eqnarray}
\begin{eqnarray}
& \Lambda_s:\hspace{5mm} &
c_1= {2\over 27} f_s,\;
c_2= -{4\over 27} f_s,\;
c_3= -{1\over 36},\;
c_4= {5\over 324} f_s,\;
c_5= -{4\over 9},\;
\label{lams33d}
\end{eqnarray}
\begin{eqnarray}
& \Lambda_c:\hspace{5mm} &
c_1= {1\over 9} f_s,\;
c_2= -{2\over 9} f_s,\;
c_3= -{1\over 24},\;
c_4= {5\over 216} f_s,\;
c_5= -{4\over 9},\;
\label{lamc33d}
\end{eqnarray}
The positive coefficient of the leading term suggests that this sum rule should be 
saturated by a negative-parity state. 
However, in the observed spectrum the ${1\over 2}^+$ state is 
lower than the ${1\over 2}^-$ state. 
Since here one does not have the freedom of varying the interpolating 
field to minimize the influence of the positive-parity term,  
the positive-parity term must be retained in a sensible fit.
This worsens its predictive ability.
Without knowing the order of the two states, it is the natural position to take to 
keep both terms.

The chiral-even sum rules coupling to both spin-3/2 and spin-1/2 states are
\begin{eqnarray}
& p_\mu p_\nu\hat{p}\; : \hspace{5mm} &
  c_1\; L^{4/27}\; E_1\; M^4
+ c_2\; b\; L^{4/27}
+ c_3\; m_s\,a\; L^{4/27}
+ c_4\; m_s\,m^2_0 a\; L^{-10/27} \; {1\over M^2}
\nonumber \\ & &
= {2\over 3}\;{\tilde{\lambda}_{3/2^-}^2 \over M^2_{3/2^-}}\;
e^{-M^2_{3/2^-}/M^2}
+ {2\over 3}\;{\tilde{\lambda}_{3/2^+}^2 \over M^2_{3/2^+}}\;
e^{-M^2_{3/2^+}/M^2}
\nonumber \\ & &
+ 16\;{\tilde{\alpha}_{1/2^+}^2 \over M^2_{1/2^+}}\; e^{-M^2_{1/2^+}/M^2}
+ 16\;{\tilde{\alpha}_{1/2^-}^2 \over M^2_{1/2^-}}\; e^{-M^2_{1/2^-}/M^2},
\label{sum33e}
\end{eqnarray}
where the coefficients are given by
\begin{eqnarray}
& N:\hspace{5mm} &
c_1={12\over 5},\;
c_2= { 1\over 3},\;
c_3= 0\;
c_4= 0,\;
\label{nuc33e}
\end{eqnarray}
\begin{eqnarray}
& \Sigma:\hspace{5mm} &
c_1={12\over 5},\;
c_2= { 1\over 3},\;
c_3= -8f_s,\;
c_4= {4\over 3}f_s,\;
\label{sig33e}
\end{eqnarray}
\begin{eqnarray}
& \Xi:\hspace{5mm} &
c_1={ 12\over 5},\;
c_2= { 1\over 3},\;
c_3= -16f_s,\;
c_4= {20\over 3}f_s,\;
\label{xi33e}
\end{eqnarray}
\begin{eqnarray}
& \Lambda_o:\hspace{5mm} &
c_1= {4\over 5},\;
c_2= {1\over 9},\;
c_3= -{8\over 3}f_s,\;
c_4= {4\over 9}f_s,\;
\label{lamo33e}
\end{eqnarray}
\begin{eqnarray}
& \Lambda_s:\hspace{5mm} &
c_1= {4\over 5},\;
c_2= {1\over 9},\;
c_3= -{8\over 3}f_s,\;
c_4= {8\over 9}f_s,\;
\label{lams33e}
\end{eqnarray}
\begin{eqnarray}
& \Lambda_c:\hspace{5mm} &
c_1= {4\over 5},\;
c_2= {1\over 9},\;
c_3= -{8\over 3}f_s,\;
c_4= {4\over 3}f_s,\;
\label{lamc33e}
\end{eqnarray}

The chiral-odd sum rules coupling to both spin-3/2 and spin-1/2 
states are
\begin{eqnarray}
& p_\mu p_\nu\; : \hspace{5mm} &
c_1\; m_s\; L^{-8/27}\; E_1\; M^4
+c_2\; a\; L^{16/27}\; E_0\; M^2
+c_3\; m^2_0 a\; L^{2/27}
\nonumber \\ & &
+c_4\; m_s \,b\; L^{-8/27}
+c_5\; a\,b\; L^{16/27}\; {1\over M^2}
+c_6\; m_s\,\kappa_v a^2\; L^{16/27}\; {1\over M^2}
\nonumber \\ & &
= - {2\over 3}\;{\tilde{\lambda}_{3/2^-}^2 \over M^2_{3/2^-}}\;
e^{-M^2_{3/2^-}/M^2}
+ {2\over 3}\;{\tilde{\lambda}_{3/2^+}^2 \over M^2_{3/2^+}}\;
e^{-M^2_{3/2^+}/M^2}
\nonumber \\ & &
- 8\;{\tilde{\alpha}_{1/2^+}^2 \over M^2_{1/2^+}}\; e^{-M^2_{1/2^+}/M^2}
+ 8\;{\tilde{\alpha}_{1/2^-}^2 \over M^2_{1/2^-}}\; e^{-M^2_{1/2^-}/M^2},
\label{sum33f}
\end{eqnarray}
where the coefficients are given by
\begin{eqnarray}
& N:\hspace{5mm} &
c_1= 0,\;
c_2= -8,\;
c_3= 8,\;
c_4= 0,\;
c_5= -1,\;
c_6= 0,\;
\label{nuc33f}
\end{eqnarray}
\begin{eqnarray}
& \Sigma:\hspace{5mm} &
c_1= 2,\;
c_2= {8\over 3}(f_s-4),\;
c_3= -{8\over 3}(f_s-4),\;
c_4= {1\over 6},\;
c_5= {1\over 3}(f_s-4),\;
c_6= {32\over 3}f_s,\;
\label{sig33f}
\end{eqnarray}
\begin{eqnarray}
& \Xi:\hspace{5mm} &
c_1= -8,\;
c_2= -{8\over 3}(4f_s-1),\;
c_3= {8\over 3}(4f_s-1),\;
c_4= -{2\over 3},\;
c_5= -{1\over 3}(4f_s-1),
\nonumber \\ & &
c_6= {16\over 3}f_s(2f_s-1),\;
\label{xi33f}
\end{eqnarray}
\begin{eqnarray}
& \Lambda_o:\hspace{5mm} &
c_1= -{2\over 3},\;
c_2= -{8\over 9}f_s,\;
c_3= {8\over 9}f_s,\;
c_4= -{1\over 18},\;
c_5= -{1\over 9}f_s,\;
c_6= 0,\;
\label{lamo33f}
\end{eqnarray}
\begin{eqnarray}
& \Lambda_s:\hspace{5mm} &
c_1= -{4\over 3},\;
c_2= -{16\over 9}f_s,\;
c_3= {16\over 9}f_s,\;
c_4= -{1\over 9},\;
c_5= -{2\over 9}f_s,\;
c_6= 0,\;
\label{lams33f}
\end{eqnarray}
\begin{eqnarray}
& \Lambda_c:\hspace{5mm} &
c_1= -2,\;
c_2= -{8\over 3}f_s,\;
c_3= {8\over 3}f_s,\;
c_4= -{1\over 6},\;
c_5= -{1\over 3}f_s,\;
c_6= 0,\;
\label{lamc33f}
\end{eqnarray}
Note that we explicitly keep the four lowest states: $3/2+, 3/2-, 1/2+, 1/2-$.
That too many unknown parameters are present in the two sum rules makes it 
impossible to isolate the different states in a straightforward Monte-Carlo analysis. 
Their usefulness only lies in serving as a consistency check 
using the parameters extracted from the other sum rules.  

\section{Analysis}
\label{analysis}

To extract physics information from the sum rules,
we employ a Monte-Carlo-based analysis procedure first introduced in~\cite{Derek96}.
The basic steps of the procedure are as follows.
Given the uncertainties in the QCD input 
parameters, randomly-selected, Gaussianly-distributed sets are generated,
from which an uncertainty distribution in the OPE can
be constructed.  Then a $\chi^2$ minimization is applied to the sum
rule by adjusting the phenomenological fit parameters.  This is done
for each QCD parameter set, resulting in distributions for
phenomenological fit parameters, from which errors are derived.
Usually, 100 such configurations are sufficient for getting stable
results. We generally select 500 or more sets which help resolve more subtle
correlations among the QCD parameters and the phenomenological fit parameters.
One distinctive advantage of the method is that the entire phase-space of the 
input QCD parameters is explored simultaneously, leading to more realistic uncertainty 
estimates in the phenomenological parameters.

In addition, the Borel window over which the two sides
of a sum rule are matched is determined by the following two criteria:
a) {\em OPE convergence} --- the highest-dimension-operators
contribute no more than 10\% to the QCD side;
b) {\em ground-state dominance} --- excited
state contributions should not exceed more than 50\% of the 
phenomenological side.
The former effectively establishes a lower limit, the latter an upper limit.
Those sum rules which do not have a valid Borel window under these 
criteria are considered unreliable.
The emphasis here is on exploring
the QCD parameter space via Monte Carlo.  The 10\%-50\% criteria are a
reasonable choice that provide a basis for quantitative analysis.
Reasonable alternatives to the criteria are automatically
explored in the Monte-Carlo analysis, as the condensate values and the
continuum threshold would change in each sample.

The QCD input parameters and their uncertainty assignments are 
given as follows.
The quark condensate in standard notation 
is taken as $a=0.52\pm0.05$ GeV$^3$, corresponding to a central value of 
$\langle\bar{u}u\rangle=-(236)^3$ MeV$^3$.
For the gluon condensate,
early estimates from charmonium~\cite{SVZ79} place it at 
$b=0.47\pm0.2$ GeV$^4$, a value commonly used in QCD sum rule 
analysis. But more recent investigations support 
much larger values~\cite{Marrow87,Bert88,Narison95,Ji95}. Here we adopt 
$b=1.2\pm0.6$ GeV$^4$ with 50\% uncertainty.
The mixed condensate parameter is placed at $m^2_0=0.72\pm0.08$ GeV$^2$.
For the four-quark condensate, there are claims of significant 
violation of the factorization hypothesis~\cite{Marrow87,Bert88,Narison95}.
Here we use $\kappa_v=2\pm 1$ and $1\leq \kappa_v \leq 4$.
The QCD scale parameter is restricted to $\Lambda_{QCD}$=0.15$\pm$0.04 GeV.
We find variations of $\Lambda_{QCD}$ have little effects on the results. 
The strange quark mass is taken as $m_s=0.15\pm 0.02$ GeV.
The value of $f_s$ has been determined in~\cite{RRY85,Bely83} and is
given by $f_s=0.83\pm0.05$ after converting to our notation by $\gamma=f_s-1$.
These uncertainties are assigned conservatively and in accordance with the
state-of-the-art in the literature. 
While some may argue that some
values are better known, others may find that the errors are
underestimated. In any event, one will learn how the uncertainties in
the QCD parameters are mapped into uncertainties in the
phenomenological fit parameters.

Before going into the numerical analysis of the sum rules,
a general discussion of the reliability of QCD sum rules is in order.
It has been argued that chirality plays an important role in determining the 
reliability of baryon QCD sum rules~\cite{Jin97}. 
One reason is that the contributions of positive- and 
negative-parity excited states partially cancel each other in the chiral-odd sum rules
and add up in the chiral-even sum rules. 
An inspection of the 11 sum rules in this work indeed exhibits this feature.
In the six chiral-odd sum rules, 
Eq.~(\ref{sum11b}) from the spin-1/2 correlator,
Eq.~(\ref{sum13b}) and Eq.~(\ref{sum13c}) from the mixed correlator, and 
Eq.~(\ref{sum33b}), Eq.~(\ref{sum33d}), and Eq.~(\ref{sum33f}) from the spin-3/2 correlator,
the positive- and negative-parity states have the opposite sign, whereas
in the chiral-even sum rules, they have the same sign.
As a result, the continuum contamination is much 
smaller in the chiral-odd sum rules than in the chiral-even ones. 
In addition, the chiral-even sum rules suffer from other relatively large uncertainties,
arising from large perturbative corrections in the OPE, 
the factorization assumption of chiral-even condensates, 
and the onset of non-factorizable condensates in relatively low dimensions~\cite{Derek90}.
So it is expected that the chiral-odd sum rules are generally more reliable than the 
chiral-even sum rules before a numerical analysis.

\subsection{Predictive Power: Three Parameter Search}
\label{res_3}

A three-parameter search is by treating all three phenomenological parameters 
(continuum threshold $w$, coupling $\lambda^2$ and baryon mass $M_B$) 
as free parameters simultaneously.
Since the QCD input parameter space is also explored simultaneously by the Monte-Carlo 
approach, a successful three-parameter search provides the best demonstration that
a QCD sum rule has predictive power.

We performed a sweeping analysis of all the 11 sum rules for each baryon member.
As expected, we found that the chiral-odd sum rules perform better than the chiral-even 
sum rules.  Only three sum rules, all chiral-odd, allow a 3-parameter search. 
The results are discussed below.

%
\begin{table}[tb]
\caption{Predictions from 3-parameter search of the chiral-odd sum rules 
Eq.~(\protect\ref{sum11b}), at the structure
$1$ from the spin-1/2 correlator. 
The Borel regions determined by OPE convergence and pole dominance are also given.
Experimental masses are taken from~\protect\cite{pdg00}.}
\label{tab_3_1}
\begin{tabular}{lcccccc}
Sum Rule & Region & $w$ & $\tilde{\lambda}_{1/2}^2$ & Mass &
Exp.\\
& (GeV) &  (GeV) & (GeV$^6$) & (GeV) & (GeV)\\ \hline
$N_{{1\over 2}+}$ ($\beta=-0.8$)
& 0.75 to 0.91  & 1.24$\pm$ .16 & 0.70$\pm$ .26 & 1.02$\pm$ .11 &
0.938 \\
$\Sigma_{{1\over 2}-}$ ($\beta=-6$)
& 1.00 to 1.37  & 1.96$\pm$ .57 & 87.2$\pm$ 53.2 & 1.54 $\pm$ .34 & 1.620
\\
$\Xi_{{1\over 2}-}$ ($\beta=-6$)
& 1.08 to 1.22 &  2.03$\pm$ .68  & 109.6$\pm$ 84.3 & 1.55$\pm$ .38 & 1.69(?)
\\
$\Lambda_{o{1\over 2}-}$ ($\beta=-6$)
& 1.08 to 1.12 &  1.72$\pm$ .46  & 60.3$\pm$ 44.4 & 1.38$\pm$ .32 & 1.670
\\
$\Lambda_{s{1\over 2}-}$ ($\beta=-6$)
& 1.18 to 1.24 &  1.78$\pm$ .20  & 41.2$\pm$ 16.3 & 1.43$\pm$ .13 & 1.405
\\
$\Lambda_{c{1\over 2}-}$ ($\beta=-6$)
& 1.10 to 1.25 &  1.80$\pm$ .23  & 49.2$\pm$ 22.5 & 1.46$\pm$ .14 & 1.405
\\
\end{tabular}
\end{table}
%
The first is Eq.~(\ref{sum11b}) from the spin-1/2 correlator.
Traditionally, it is a controversial sum rule regarding its performance.
For a long time, its partner, the chiral-even Eq.~(\ref{sum11a}), 
was identified as the better sum rule with the value of $\beta$ taken as $-1$.
It was proven otherwise in the Monte-Carlo analysis of~\cite{Derek96}.
Moreover, for the chiral-odd sum rule, the optimal mixing
for the nucleon was determined to be $\beta=-0.8$ rather than $-1$.
This result for the nucleon is reproduced in this work.
On the other hand, in our analysis of the $\Sigma$, $\Xi$, and $\Lambda$ states, we 
could not find a solution when we assume the sum rule is saturated 
by a positive-parity state while tuning $\beta$ in a wide range.
Instead we found that the 3-parameter search is successful 
when we assume the sum rule is dominated by a negative-parity state for $\beta>-2$.
The results for $\beta=-6$ are given in Table~\protect\ref{tab_3_1}.
The masses and thresholds are quite robust, 
insensitive to variations in $\beta$ in the range of $-2$ to $-20$.
The only thing that changes with $\beta$ is the coupling $\lambda^2$ as it should be 
$\beta$ dependent.
So we conclude that this sum rule predicts a positive-parity state
in the non-strange channel with $\beta=-0.8$, 
but the same sum rule predicts negative-parity states in the strange channels
with a large negative value of $\beta$ which 
means a large negative component of $\chi_2$ in the interpolating field in 
Eq.~(\ref{nucleon_beta}).
The uncertainties were derived from 500 sets of QCD parameters by Monte-Carlo.
The predictions compare favorably with experiments.
In the case of $\Xi_{{1\over 2}-}$, the QCD sum rule predicts a mass of about 1.60 GeV.
The Particle Data Group~\cite{pdg00} lists two 3-star $\Xi$ states with 
unknown spin-parity, 1690 MeV and 1950 MeV.  Our prediction favors the lower one.
The three Lambda types give roughly the same masses within uncertainties.

The above is an example of how a sum resolves the parity of a state.
Recall that a baryon interpolating field couples to both 
positive- and negative-parity states.
That is why two terms on the phenomenological side are explicitly kept, 
one of each parity. 
But keeping two terms increases the number phenomenological parameters from 3 to 5,
which is too many in practice.
Rarely a QCD rum rule has enough information to yield a successful 5-parameter search.
So the standard practice is to use a single pole plus continuum model with 3 parameters
to maintain a reasonable level of predictive power.
But which of the two terms dominates a QCD sum rule
depends on the interpolating field.
Tuning $\beta$ changes the makeup of the interpolating field, therefore alters 
the coupling strength to the states. The advantage of having an adjustable 
interpolating field is that one could vary it to have maximal coupling to a state of definite parity. 
As a consistency check,
we kept both terms and search the couplings $\tilde{\lambda}_{1/2^+}^2$, 
$\tilde{\lambda}_{1/2^-}^2$, and threshold $w$ simultaneously by setting the masses 
for $N_{1/2+}$ and $N_{1/2-}$ at the experimental values. 
The result is that the coupling $\tilde{\lambda}_{1/2^+}^2$
comes out several orders of magnitude greater than $\tilde{\lambda}_{1/2^-}^2$, 
showing that the sum rule is indeed dominated by the $N_{1/2+}$ state.
The same check shows that for the strange baryons, the negative-parity states dominate.

The next chiral-odd sum rule that permits a 3-parameter search is 
Eq.~(\ref{sum13c}) at the tensor structure $\gamma_\mu p_\nu \hat{p}$
from the mixed correlator.  
The results are shown in Table~\protect\ref{tab_3_3}.
This is probably the best sum rule in predicting the $J^P={1\over /2}^+$ states. 
The Borel windows are relatively wide.
The predicted masses are reasonably close to the experimental values. 
And the results are stable, and insensitive 
to the value of $\beta$ as long as $\beta > 0$. 
The sum rules for $\Lambda_{o{1\over 2}+}$ and  $\Lambda_{c{1\over 2}+}$ give 
nearly identical results and the masses slightly overestimates the experimental numbers. 
We could not find a solution for the singlet $\Lambda_{s{1\over 2}+}$.
The uncertainty on the masses from the Monte-Carlo analysis is on the order of 100 MeV.

The other chiral-odd sum rule from the mixed correlator, 
Eq.~(\ref{sum13b}) at the $\gamma_\mu \gamma_\nu$ structure,
did not allow a 3-parameter search under the established fitting policy.
Our analysis showed that the continuum contributions in these sum rules are 
relatively large (over 50\% contribution), thus no valid Borel windows exist. 
However, if we relax the pole-dominance criterion and allow continuum contributions 
to exceed 50\%, we could obtain reasonable and stable 3-parameter search results for 
$J^P={1\over /2}^+$. The results will not be shown because of this compromise.

%
\begin{table}[tb]
\caption{Predictions from 3-parameter search of the chiral-odd sum rules,
Eq.~(\protect\ref{sum13c}), at the tensor structure
$\gamma_\mu p_\nu \hat{p}$ from the mixed correlator.}
\label{tab_3_3}
\begin{tabular}{ccccccc}
Sum Rule & Region & $w$ & $\tilde{\lambda}_{1/2} \tilde{\alpha}_{1/2}$ & Mass &
Exp.\\
& (GeV) &  (GeV) & (GeV$^6$) & (GeV) & (GeV)\\ \hline
$N_{{1\over 2}+} \, (\beta=+1.0)$
& 1.06 to 1.46  & 1.31$\pm$ .22 & 1.13$\pm$ .53 & 1.06$\pm$ .11 &
0.938 \\
$\Sigma_{{1\over 2}+} \, (\beta=+1.0) $
& 1.12 to 1.53  & 1.48$\pm$ .23 & 1.62$\pm$ 0.68 & 1.16 $\pm$ .12 & 1.193
\\
$\Xi_{{1\over 2}+} \, (\beta=+1.0)$
& 1.35 to 1.80 & 1.69$\pm$ .27  & 2.61$\pm$ 1.20 & 1.32$\pm$ .14 & 1.318
\\
$\Lambda_{o{1\over 2}+} \, (\beta=+1.0)$
& 1.27 to 1.74 &  1.53$\pm$ .24  & 0.66$\pm$ 0.28 & 1.23$\pm$ .12 & 1.116
\\
$\Lambda_{c{1\over 2}+} \, (\beta=+1.0)$
& 1.28 to 1.72 &  1.53$\pm$ .24  & 0.66$\pm$ 0.28 & 1.23$\pm$ .12 & 1.116
\\
\end{tabular}
\end{table}
%

Next we examine the chiral-odd sum rule of 
Eq.~(\protect\ref{sum33b}) at the structure $g_{\mu\nu}$ 
from the spin-3/2 correlator.
which couples only to spin-3/2 baryon states. 
It allows a 3-parameter search and is saturated by $J^P={3\over /2}^-$ states. 
One feature here is that there is no tunable parameter in the spin-3/2 correlator,  
which leaves the Monte-Carlo analysis quite straightforward. The results are given
in Table~\protect\ref{tab_3_4}. 
If we explore the contribution from each term of the OPE, 
we can easily see by comparing $\Sigma$ and 
$\Xi$ that the prominent difference is on the leading term, 
which originates from flavor-symmetry breaking via the strange quark mass $m_s$. 
For $\Sigma$, this term acts to decrease the mass of $\Sigma_{{3\over 2}-}$, 
while $\Xi_{{3\over 2}-}$ it acts to increase its mass. 

%
\begin{table}[tb]
\caption{Predictions from 3-parameter search of the chiral-odd sum rules,
Eq.~(\protect\ref{sum33b}) at the structure $g_{\mu\nu}$ from the spin-3/2 correlator.}
\label{tab_3_4}
\begin{tabular}{ccccccc}
Sum Rule & Region & $w$ & $\tilde{\lambda}_{3/2}^2$ & Mass &
Exp.\\
& (GeV) &  (GeV) & (GeV$^6$) & (GeV) & (GeV)\\ \hline
$N_{{3\over 2}-}$
& 0.95 to 1.17  & 1.65$\pm$ .24 & 27.6$\pm$ 11.8 & 1.44$\pm$ .13 &
1.520 \\
$\Sigma_{{3\over 2}-}$
& 1.29 to 1.36  & 1.91$\pm$ .25 & 46.6$\pm$ 20.1 & 1.69 $\pm$ .14 & 1.580
\\
$\Xi_{{3\over 2}-}$
& 1.30 to 1.39 & 2.19$\pm$ .27  & 84.8$\pm$ 42.9 & 1.84$\pm$ .16 & 1.820
\\
$\Lambda_{o{3\over 2}-}$
& 1.48 to 1.55 &  2.36$\pm$ .35  & 12.7$\pm$ 7.4 & 2.00$\pm$ .21 & 1.690
\\
$\Lambda_{s{3\over 2}-}$
& 1.30 to 1.39 &  2.12$\pm$ .28  & 16.8$\pm$ 8.7 & 1.80$\pm$ .16 & 1.690
\\
$\Lambda_{c{3\over 2}-}$
& 1.22 to 1.32 &  2.01$\pm$ .25  & 19.8$\pm$ 9.3 & 1.71$\pm$ .15 & 1.520
\\
\end{tabular}
\end{table}
%

\subsection{Predictive Power: Multiple Sum Rule Search}
\label{resmm}

Besides the 3-parameter search for a single sum rule, we made further efforts
to explore the predictive ability of the QCD sum rules.
One possibility that is afforded by the Monte-Carlo method 
is to perform a simultaneous search of multiple sum rules,
each having an independent Borel window and continuum threshold.
The benefit is that more information is introduced on the OPE side 
and at the same time the number of parameters 
that are shared across the multiple sum rules is reduced.
This may lead to increased stability and smaller errors in the output.
In principle, it should be done with those sum rules that come from the same correlator. 
The only successful multiple sum rule search is from the sum rules 
of Eq.~(\ref{sum13b}) and Eq.~(\ref{sum13b}) from the mixed correlator, 
as shown in Table~\protect\ref{tabMsearch}.
It is essentially a 5-parameter, 2-sum rule search for the $J^P={1\over 2}^+$ states.
The results improve upon those in Table~\ref{tab_3_3}.
In the case of flavor-singlet $\Lambda$, it predicts a mass of around 2.6 GeV. 
So QCD sum rules predicts the correct ordering of flavor-singlet Lambda with 
${1\over 2}^-$ state of 1.405 GeV, and ${1\over 2}^+$ state of much higher mass. 
This relatively high value for the $J^P={1\over 2}^+$ flavor-singlet Lambda is yet to 
be identified by experiment.
The octet $\Lambda$ and the common $\Lambda$ interpolating fields produce comparable 
results.

%
\begin{table}[tb]
\caption{Results from a 5-parameter, simultaneous search of the two chiral-odd sum rules of
Eq.(~\protect\ref{sum13b}) at the structure $\gamma_\mu\gamma_\nu $ (labeled as 1) and 
Eq.(~\protect\ref{sum13c}) at the structure $\gamma_\mu p_\nu \hat p$ (labeled as 2)
from the mixed correlator.}
\label{tabMsearch}
\begin{tabular}{cccccccc}
Sum Rule & Region1    & $w$1 & Region2 & $w$2
         & $\tilde{\lambda}_{1/2+} \tilde{\alpha}_{1/2+} $ & Mass & Exp. \\
         & (GeV)  & (GeV) &(GeV) & (GeV) & (GeV$^6$) & (GeV) & (GeV)     \\ \hline
$N_{{1\over 2}+}$ ($\beta=0$)  & 0.69 to 1.12 & 1.44 $\pm$.15 & 0.80 to 1.52  &  1.35 $\pm$ .16 
              & 1.07$\pm$.25 & 1.02$\pm$ .07 & 0.938\\
$\Sigma_{{1\over 2}+}$ ($\beta=-1$) &0.75 to 1.24 &1.67 $\pm$ .25 & 0.88 to 1.68 & 1.58 $\pm$ .25 
                     & 1.72 $\pm$ .66 & 1.17$\pm$ .11 & 1.193 \\
$\Xi_{{1\over 2}+}$ ($\beta=-1$)  & 0.97 to 1.34 &1.89 $\pm$ .32 & 1.10 to 1.81 & 1.80 $\pm$ .30 
                   & 2.70 $\pm$ 1.23 & 1.32$\pm$ .14 & 1.318\\
$\Lambda_{o{1\over 2}+}$ ($\beta=-1$) & 0.74 to 1.24 &1.67 $\pm$ .25 & 0.88 to 1.66 & 1.58 $\pm$ .25 
                        & 0.57 $\pm$ .22 & 1.17 $\pm$ .11 & 1.116\\
$\Lambda_{s{1\over 2}+}$ ($\beta=-1$) & 1.41 to 1.98 & 3.46 $\pm$ .28 & 1.64 to 2.45 & 3.55 $\pm$ .39 
                        & 1.71 $\pm$ .72 & 2.65 $\pm$ .21 & ?\\
$\Lambda_{c{1\over 2}+}$ ($\beta=-1$)  &0.80 to 1.06 &1.53 $\pm$ .45 & 0.86 to 1.21 & 1.27 $\pm$ .08 
                        & 0.38 $\pm$ .05 & 1.09 $\pm$ .03 & 1.116
\end{tabular}
\end{table}
%

\subsection{Two Parameter Search Results}
\label{res11}

Absence of a 3-parameter search, the alternative is to perform 
a 2-parameter search by fixing one of the parameters to a certain value.
In our analysis we use the following procedure.
We start by fixing the baryon mass at the experimental value 
and search for the continuum threshold and coupling 
simultaneously.  Then we fix the continuum threshold to the returned value and search 
for the mass and the coupling.  To make sure the consistency of above two searches, 
a third search for the threshold and the mass is conducted by fixing the coupling 
to the returned value.  
Fitting results of the three parameters are accepted if above three searches 
yield consistent results.
In the absence of such consistency, 
we reduce to giving results from just fixing the continuum 
threshold and searching for the mass and the coupling, which is conventionally done.

Results from such two-parameter searches should be treated with caution because they 
are not truly predicted in the sense established earlier.
It is no secret that for most sum rules it is possible to produce the desired mass 
by fixing the continuum threshold to a favorable value and adjusting the Borel window 
accordingly.
In such cases, it is possible that the continuum contribution is too large, and 
the sum rule is analyzed deep in the perturbative regime, away from the cross-over region 
between perturbative and non-perturbative where the QCD sum rule is based.

As an example, 
we turn our attention to the chiral-even sum rule of Eq.~(\ref{sum11a}) 
at the $\hat p$ structure from the spin-1/2 correlator.
This sum rule has been traditionally identified as the more reliable of the 
two sum rules from the spin-1/2 correlator.
However, it was shown by Leinweber~\cite{Derek96}
that a native 3-parameter search with the conventional value of $\beta=-1$ 
gives a prediction of the nucleon mass around 450 MeV, which is very different 
from the observed value. Closer examination revealed that the sum rule suffers from large 
uncertainties in the OPE and a lack of a valid Borel window, 
so a three-parameter search is unstable.
However, we found that a two-parameter search could produce sensible results 
for this sum rule, as shown in Table~\protect\ref{tab11a}.
We set the beta value to the traditional value of $\beta=-1$.
The results compare favorably with experiments.
Again it favors a high value of around 2.6 GeV for the $J^P={1\over 2}^+$, 
flavor-singlet $\Lambda$. No solution could be found for the common $\Lambda$.

Another sum rule that can produce sensible  two-parameter search results is 
the chiral-odd sum rule of Eq.~(\ref{sum13b})
at the structure $\gamma_\mu \gamma_\nu$, as shown in Table~t\ref{tab13b}.
The results have broad agreement with those in Table~\ref{tab11a}.
The fits are very stable and have smaller than usual uncertainties in the parameters.

%
\begin{table}[tb]
\caption{Two-parameter search results from the chiral-even sum rules of
Eq.~(\protect\ref{sum11a}) at the structure $\hat{p}$ from the spin-1/2 correlator.}
\label{tab11a}
\begin{tabular}{ccccccc}
Sum Rule & Region & $w$ & $\tilde{\lambda}_{1/2}^2$ & Mass & Exp.\\
& (GeV) &  (GeV) & (GeV$^6$) & (GeV) & (GeV)\\ \hline
$N_{{1\over 2}+}$ $(\beta=-1)$
& 0.77 to 1.28  & 1.88$\pm$ .17 & 1.97$\pm$ .18 & 0.90$\pm$ .10 & 0.938 \\
$\Sigma_{{1\over 2}+}$ $(\beta=-1)$
& 0.79 to 1.40  & 2.14$\pm$ .10 & 3.45$\pm$ .25 & 1.15 $\pm$ .17 & 1.193 \\
$\Xi_{{1\over 2}+}$ $(\beta=-1)$
& 0.74 to 1.43 &  2.27$\pm$ .15  & 4.50$\pm$ .94 & 1.21$\pm$ .11 & 1.318 \\
$\Lambda_{o{1\over 2}+}$ $(\beta=-1)$
& 0.73 to 1.35 & 2.07$\pm$ .16    & 2.76$\pm$ .16 & 1.08 $\pm$ .09& 1.116 \\
$\Lambda_{s{1\over 2}+}$ $(\beta=-1)$
& 1.19 to 1.98 & 3.16 $\pm$.33 &  51.5$\pm$19.0 & 2.64 $\pm$ .16 &   
\end{tabular}
\end{table}
%

%
\begin{table}[tb]
\caption{Two-parameter search results from the chiral-odd sum rules of
Eq.~(\protect\ref{sum13b}) at the structure
$\gamma_\mu\gamma_\nu $ from the mixed correlator.}
\label{tab13b}
\begin{tabular}{ccccccc}
Sum Rule & Region  & $w$ &
$\tilde{\lambda}_{1/2}\tilde{\alpha}_{1/2}$ & Mass & Exp.\\
& (GeV) & (GeV) & (GeV$^6$) & (GeV) & (GeV)\\ \hline
$N_{{1\over 2}+}$ $(\beta=0)$ 
& 0.68 to 1.03 & 1.31$\pm$ .03 & 0.81$\pm$ .90 & 0.94 $\pm$ .02 & 0.938 \\
$\Sigma_{{1\over 2}+}$  $(\beta=-1)$
& 0.756 to 1.28  & 1.72 $\pm$ .07      & 1.81$\pm$ .30 & 1.18$\pm$ .04& 1.193 \\
$\Xi_{{1\over 2}+}$ $(\beta=-1)$
& 0.976 to 1.36  & 1.88 $\pm$ .06     & 2.55$\pm$ .42 & 1.31 $\pm$ .03  & 1.318 \\
$\Lambda_{o{1\over 2}+}$ $(\beta=-1)$
& 0.75 to 1.19  & 1.60 $\pm$ .08      & 0.49$\pm$ .09 & 1.11 $\pm$ .04 & 1.116 \\
$\Lambda_{s{1\over 2}+}$ $(\beta=-1)$
& 1.42 to 1.96  & 3.30 $\pm$ .39      & 1.38$\pm$ .67 & 2.67 $\pm$ .19 &       \\
$\Lambda_{c{1\over 2}+}$ $(\beta=-1)$
& 0.80 to 1.03  & 1.45 $\pm$ .05      & 0.34$\pm$ .05 & 1.12 $\pm$ .02 & 1.116 \\
\end{tabular}
\end{table}
%

The chiral-even sum rule of Eq.~(\ref{sum33c}) at the structure $\gamma_\mu p_\nu +\gamma_\nu p_\mu$
and the chiral-odd sum rule of Eq.~(\ref{sum33d}) at the structure 
$\gamma_\mu\gamma_\nu + {1\over 3}\, g_{\mu \nu}$
in the spin-3/2 correlator couple only to $J=\pm 1/2$ baryon states. 
Numerical analysis confirms that the chiral-even one is a poor sum rule with nothing produced. 
The chiral-odd one is dominated by both positive-parity and negative-parity states.
The Monte Carlo fit is unable to be performed without the positive-parity term. 
We tested this by fitting the couplings $\tilde{\alpha}_{1/2^+}^2$, 
$\tilde{\alpha}_{1/2^-}^2$, and 
threshold $w$ simultaneously and fixing the masses of $B_{1/2+}$ and $B_{1/2-}$ 
at experimental values.  The two coupling constants came out very close to each other, 
which means the positive- and negative-parity states are equally important in this sum rule.
We could obtain reasonable results for negative-parity states by fixing the 
positive-parity masses at their observed values. 
But the continuum contributions were found to be too large in the fits.
Therefore we decided not to include the results here.

The last two sum rules of Eq.~(\ref{sum33e}) and Eq.~(\ref{sum33f}) involve a mixture of 
${3\over 2}^-$ and ${1\over 2}^\pm$ states,
which makes them not suitable for a Monte-Carlo search in their own right 
due to the numerous unknown parameters.
Their utility lies in serving as a consistency check of the other sum rules.
For example, when we fixed the parameters of the ${1\over 2}^\pm$ states in 
the chiral-odd sum rule of Eq.~(\ref{sum33f}) to the values determined from 
the other sum rules, and searched for the continuum threshold and the 
${3\over 2}^-$ parameters,
we obtained results that are consistent with those in Table~\ref{tab_3_4}.

\section{Summary and Conclusion}
\label{con}

We have performed a comprehensive study of all 11 QCD sum rules using
general interpolating fields for spin-1/2 and spin-3/2 baryon states.
They are derived from three types of correlation functions:
2 from the spin-1/2 correlator, 3 from the mixed correlator, and 6 from the spin-3/2 correlator. 
In terms of the chirality of the QCD operators in the sum rules, 
the breakdown is that 
1+2+3=6 are chiral-odd (which contain dimension-odd QCD condensates only), 
and 1+1+3=5 are chiral-even (which contain dimension-even QCD condensates only).
Our results lend support to the general statement that chiral-odd sum rules are 
better than chiral-even sum rules as far as
baryon two-point functions are concerned~\cite{Jin97}.

The sum rules contain information on the masses of $N$, $\Sigma$, $\Xi$ and $\Lambda$
states with spin-parity $J^P={1\over 2}^\pm$ and $J^P={3\over 2}^\pm$.
A Monte-Carlo based method is used to assess whether a given QCD sum rule has 
good predictive ability which we define as allowing a 3-parameter search 
in a $\chi^2$ minimization.
We found that 3 sum rules, all chiral-odd,
Eq.~(\ref{sum11b}) from the spin-1/2 correlator,
Eq.~(\ref{sum13c}) from the mixed correlator, and 
Eq.~(\ref{sum33b}) from the spin-3/2 correlator, have good predictive ability.
They give predictions for the ${1\over 2}^-$ states in Table~\ref{tab_3_1},
for ${1\over 2}^+$ states in Table~\ref{tab_3_3},
and for ${3\over 2}^-$ states in Table~\ref{tab_3_4}.
Furthermore, the Monte-Carlo method affords the ability to fit multiple sum rules 
at the same time. 
A multiple search of the two chiral-odd sum rules from the mixed correlator yields 
even better results for ${1\over 2}^+$ states as given in Table~\ref{tabMsearch}.

The chiral-odd sum rules from the mixed operator produce 
sum rules that have better OPE convergence which in turn lead to robust predictions.
The results bode well for the use of mixed correlators in future studies.
We also performed conventional, two-parameter searches by fixing one of the parameters.
We can reproduce the results in the literature, and obtain sensible new results. But
such results have certain bias built-in from the start and should be taken with caution.

The predictions compare favorably with the observed values, with an uncertainty 
on the order of 100 MeV. The couplings come as by-products which are useful in 
the calculation of matrix elements because they enter as normalization.
In the case of flavor-singlet $\Lambda$ with ${1\over 2}^+$, 
the QCD sum rule approach predicts a high value of around 2.6 GeV.
In the case of $\Xi$, 
the QCD sum rule approach favors the state with 1690 MeV as having ${1\over 2}^+$, 
over the heavier state of 1950 MeV, both of which have unknown spin-parity 
assignment in the particle data table.
Taken together, the results show that the QCD sum rule approach has predictive power 
for the masses of the excited baryons considered in this work, but that 
such predictive power must be examined on a sum rule by sum rule basis with 
rigorous analysis.

One drawback in the current approach is that states with both parities
contribute in the same sum rule for a baryon. 
Although sometimes one can saturate a sum rule
with a certain parity, by using a general interpolating field with a 
adjustable parameter (like $\beta$), it is not always feasible.
For example, no information could be extracted for the ${3\over 2}^+$ states
because no sum rules from the interpolating fields considered in this work 
are dominated by them.
It is desirable to perform a parity separation, as proposed in Ref.~\cite{Jido96}, 
which can be achieved by replacing the time-ordering operator $T$
in the correlation function in Eq.~(\ref{cf2pt}) with $x_0>0$,
and constructing sum rules in the complex $p_0$-space in the rest frame ($\vec{p}=0$).
Work that applies the parity projection method to the sum rules 
in this work and that analyzes them with the Monte-Carlo procedure 
will be reported elsewhere~\cite{Lee02}.

\acknowledgements
This work was supported in part by U.S. DOE under Grant DE-FG03-93DR-40774.
We are grateful to D.B. Leinweber for sharing an original version
of his Monte-Carlo analysis program.

\appendix
\section{Master Formulas used in the Derivations}
\label{appendix}

In this appendix we collect the master formulas used in the derivation of the QCD Sum Rules.
They are functions of the quark propagators, obtained by contracting out pairs of time-ordered quark-field operators 
in the two-point correlation function in Eq.~(\ref{cf2pt}).
The same master formulas are used in lattice QCD calculations where the fully-interacting 
quark propagators are generated numerically by Monte-Carlo, instead of the analytical ones 
used here as in Eq.~(\ref{qprop}).

\subsection{Spin-1/2 Correlators}

Using the spin-1/2 interpolating field of Eq.~(\ref{nucleon_beta}) for the nucleon 
(with $uud$ quark content) , we find
\begin{eqnarray}
& &
\langle\Omega\,|\, T\{\eta^{N}_{\scriptscriptstyle 1/2}(x)\,
\bar{\eta}^{N}_{\scriptscriptstyle 1/2}(0)\,|\,\Omega\rangle
=-4\epsilon^{abc}\epsilon^{a^\prime b^\prime c^\prime} \{\;
\nonumber \\ & &
+S^{aa^\prime}_u \gamma_5 C {S^{cc^\prime}_d}^T C  \gamma_5 S^{bb^\prime}_u
+S^{aa^\prime}_u \mbox{Tr}(C {S^{cc^\prime}_d}^T C \gamma_5 S^{bb^\prime}_u \gamma_5)
\nonumber \\ & &
+\beta\gamma_5  S^{aa^\prime}_u \gamma_5 C {S^{cc^\prime}_d}^T C S^{bb^\prime}_u
+\beta \gamma_5  S^{aa^\prime}_u \mbox{Tr}( C {S^{cc^\prime}_u}^T C S^{bb^\prime}_d \gamma_5)
\nonumber \\ & &
+\beta  S^{aa^\prime}_u C {S^{cc^\prime}_d}^T C \gamma_5 S^{bb^\prime}_u \gamma_5
+\beta S^{aa^\prime}_u \gamma_5 \mbox{Tr}( C {S^{cc^\prime}_u}^T C \gamma_5 S^{bb^\prime}_d)
\nonumber \\ & &
+\beta^2 \gamma_5  S^{aa^\prime}_u C {S^{cc^\prime}_d}^T C S^{bb^\prime}_u \gamma_5
+\beta^2 \gamma_5 S^{aa^\prime}_u \gamma_5 \mbox{Tr}( C {S^{cc^\prime}_d}^T C S^{bb^\prime}_u)
\;\}.
\label{master11}
\end{eqnarray}
The master formula for $\Sigma$ (with $uus$ quark content) can be readily obtained by replacing the 
$d$ quark with a $s$ quark.
Similarly, the master formula for $\Xi$ (with $ssu$ quark content) can be obtained by a two-step 
replacement in Eq.~(\ref{master11}): $u$ replaced with $s$, followed by $d$ with $u$.
The master formulas from the three $\Lambda$ (with $uds$ quark content) 
interpolating fields have more complicated expressions, for the octet $\Lambda$:
\begin{eqnarray}
& &
\langle\Omega\,|\, T\{\;\eta^{\Lambda_o}_{\scriptscriptstyle 1/2}(x)\,
\bar{\eta}^{\Lambda_o}_{\scriptscriptstyle 1/2}(0)\;\}\,|\,\Omega\rangle
=-{2\over 3}\epsilon^{abc}\epsilon^{a^\prime b^\prime c^\prime} \{\;
\nonumber \\ & &
+2S^{aa^\prime}_s \gamma_5 C {S^{cc^\prime}_u}^T C  \gamma_5 S^{bb^\prime}_d
+2S^{aa^\prime}_s \gamma_5 C {S^{cc^\prime}_d}^T C  \gamma_5 S^{bb^\prime}_u
+4S^{aa^\prime}_s \mbox{Tr}(C {S^{cc^\prime}_u}^T C \gamma_5 S^{bb^\prime}_d \gamma_5)
\nonumber \\ & &
+2S^{aa^\prime}_d \gamma_5 C {S^{cc^\prime}_u}^T C  \gamma_5 S^{bb^\prime}_s
-S^{aa^\prime}_d \gamma_5 C {S^{cc^\prime}_s}^T C  \gamma_5 S^{bb^\prime}_u
+S^{aa^\prime}_d \mbox{Tr}(C {S^{cc^\prime}_s}^T C \gamma_5 S^{bb^\prime}_u \gamma_5)
\nonumber \\ & &
+2S^{aa^\prime}_u \gamma_5 C {S^{cc^\prime}_d}^T C  \gamma_5 S^{bb^\prime}_s
-S^{aa^\prime}_u \gamma_5 C {S^{cc^\prime}_s}^T C  \gamma_5 S^{bb^\prime}_d
+S^{aa^\prime}_u \mbox{Tr}(C {S^{cc^\prime}_s}^T C \gamma_5 S^{bb^\prime}_d \gamma_5)
\nonumber \\ & &
+2\beta S^{aa^\prime}_s C {S^{cc^\prime}_u}^T C  \gamma_5 S^{bb^\prime}_d \gamma_5
+2\beta S^{aa^\prime}_s C {S^{cc^\prime}_d}^T C  \gamma_5 S^{bb^\prime}_u \gamma_5
+4\beta S^{aa^\prime}_s \gamma_5 \mbox{Tr}(C {S^{cc^\prime}_d}^T C \gamma_5 S^{bb^\prime}_u)
\nonumber \\ & &
+2\beta S^{aa^\prime}_d C {S^{cc^\prime}_u}^T C  \gamma_5 S^{bb^\prime}_s \gamma_5
-\beta S^{aa^\prime}_d C {S^{cc^\prime}_s}^T C  \gamma_5 S^{bb^\prime}_u \gamma_5 
+\beta S^{aa^\prime}_d \gamma_5 \mbox{Tr}(C {S^{cc^\prime}_s}^T C \gamma_5 S^{bb^\prime}_u)
\nonumber \\ & &
+2\beta S^{aa^\prime}_u C {S^{cc^\prime}_d}^T C  \gamma_5 S^{bb^\prime}_s \gamma_5
-\beta S^{aa^\prime}_u C {S^{cc^\prime}_s}^T C  \gamma_5 S^{bb^\prime}_d \gamma_5
+\beta S^{aa^\prime}_u \gamma_5 \mbox{Tr}(C {S^{cc^\prime}_s}^T C \gamma_5 S^{bb^\prime}_d)
\nonumber \\ & &
+2\beta \gamma_5 S^{aa^\prime}_s \gamma_5 C {S^{cc^\prime}_u}^T C S^{bb^\prime}_d
+2\beta \gamma_5 S^{aa^\prime}_s \gamma_5 C {S^{cc^\prime}_d}^T C S^{bb^\prime}_u
+4\beta \gamma_5 S^{aa^\prime}_s \mbox{Tr}(C {S^{cc^\prime}_d}^T C S^{bb^\prime}_u \gamma_5)
\nonumber \\ & &
+2\beta \gamma_5 S^{aa^\prime}_d \gamma_5 C {S^{cc^\prime}_u}^T C S^{bb^\prime}_s
-\beta \gamma_5 S^{aa^\prime}_d \gamma_5 C {S^{cc^\prime}_s}^T C S^{bb^\prime}_u
+\beta \gamma_5 S^{aa^\prime}_d \mbox{Tr}(C {S^{cc^\prime}_s}^T C S^{bb^\prime}_u \gamma_5)
\nonumber \\ & &
+2\beta \gamma_5 S^{aa^\prime}_u \gamma_5 C {S^{cc^\prime}_d}^T C S^{bb^\prime}_s
-\beta \gamma_5 S^{aa^\prime}_u \gamma_5 C {S^{cc^\prime}_s}^T C S^{bb^\prime}_d
+\beta \gamma_5 S^{aa^\prime}_u \mbox{Tr}(C {S^{cc^\prime}_s}^T C S^{bb^\prime}_d \gamma_5)
\nonumber \\ & &
+2\beta^2 \gamma_5 S^{aa^\prime}_s C {S^{cc^\prime}_u}^T C S^{bb^\prime}_d \gamma_5
+2\beta^2 \gamma_5 S^{aa^\prime}_s C {S^{cc^\prime}_d}^T C S^{bb^\prime}_u \gamma_5
+4\beta^2 \gamma_5 S^{aa^\prime}_s \gamma_5 \mbox{Tr}(C {S^{cc^\prime}_u}^T C S^{bb^\prime}_d)
\nonumber \\ & &
+2\beta^2 \gamma_5 S^{aa^\prime}_d C {S^{cc^\prime}_u}^T C  S^{bb^\prime}_s \gamma_5
-\beta^2 \gamma_5 S^{aa^\prime}_d C {S^{cc^\prime}_s}^T C S^{bb^\prime}_u \gamma_5
+\beta^2 \gamma_5 S^{aa^\prime}_d \gamma_5 \mbox{Tr}(C {S^{cc^\prime}_s}^T C S^{bb^\prime}_u)
\nonumber \\ & &
+2\beta^2 \gamma_5 S^{aa^\prime}_u C {S^{cc^\prime}_d}^T C S^{bb^\prime}_s \gamma_5
-\beta^2 \gamma_5 S^{aa^\prime}_u C {S^{cc^\prime}_s}^T C S^{bb^\prime}_d \gamma_5
+\beta^2 \gamma_5 S^{aa^\prime}_u \gamma_5 \mbox{Tr}(C {S^{cc^\prime}_s}^T C S^{bb^\prime}_d)\;\},
\label{master11_lambda_o}
\end{eqnarray}
for the singlet $\Lambda$:
\begin{eqnarray}
& &
\langle\Omega\,|\, T\{\;\eta^{\Lambda_s}_{\scriptscriptstyle 1/2}(x)\,
\bar{\eta}^{\Lambda_s}_{\scriptscriptstyle 1/2}(0)\;\}\,|\,\Omega\rangle
=-{4\over 3}\epsilon^{abc}\epsilon^{a^\prime b^\prime c^\prime} \{\;
\nonumber \\ & &
-S^{aa^\prime}_s \gamma_5 C {S^{cc^\prime}_u}^T C  \gamma_5 S^{bb^\prime}_d
-S^{aa^\prime}_s \gamma_5 C {S^{cc^\prime}_d}^T C  \gamma_5 S^{bb^\prime}_u
+S^{aa^\prime}_s \mbox{Tr}(C {S^{cc^\prime}_d}^T C \gamma_5 S^{bb^\prime}_u \gamma_5)
\nonumber \\ & &
-S^{aa^\prime}_d \gamma_5 C {S^{cc^\prime}_u}^T C  \gamma_5 S^{bb^\prime}_s
-S^{aa^\prime}_d \gamma_5 C {S^{cc^\prime}_s}^T C  \gamma_5 S^{bb^\prime}_u
+S^{aa^\prime}_d \mbox{Tr}(C {S^{cc^\prime}_s}^T C \gamma_5 S^{bb^\prime}_u \gamma_5)
\nonumber \\ & &
-S^{aa^\prime}_u \gamma_5 C {S^{cc^\prime}_d}^T C  \gamma_5 S^{bb^\prime}_s
-S^{aa^\prime}_u \gamma_5 C {S^{cc^\prime}_s}^T C \gamma_5 S^{bb^\prime}_d
+S^{aa^\prime}_u \mbox{Tr}(C {S^{cc^\prime}_s}^T C \gamma_5 S^{bb^\prime}_d \gamma_5)
\nonumber \\ & &
-\beta S^{aa^\prime}_s C {S^{cc^\prime}_u}^T C  \gamma_5 S^{bb^\prime}_d \gamma_5
-\beta S^{aa^\prime}_s C {S^{cc^\prime}_d}^T C  \gamma_5 S^{bb^\prime}_u \gamma_5
+\beta S^{aa^\prime}_s \gamma_5 \mbox{Tr}(C {S^{cc^\prime}_d}^T C \gamma_5 S^{bb^\prime}_u)
\nonumber \\ & &
-\beta S^{aa^\prime}_d C {S^{cc^\prime}_u}^T C  \gamma_5 S^{bb^\prime}_s \gamma_5
-\beta S^{aa^\prime}_d C {S^{cc^\prime}_s}^T C  \gamma_5 S^{bb^\prime}_u \gamma_5 
+\beta S^{aa^\prime}_d \gamma_5 \mbox{Tr}(C {S^{cc^\prime}_s}^T C \gamma_5 S^{bb^\prime}_u)
\nonumber \\ & &
-\beta S^{aa^\prime}_u C {S^{cc^\prime}_d}^T C  \gamma_5 S^{bb^\prime}_s \gamma_5
-\beta S^{aa^\prime}_u C {S^{cc^\prime}_s}^T C  \gamma_5 S^{bb^\prime}_d \gamma_5
+\beta S^{aa^\prime}_u \gamma_5 \mbox{Tr}(C {S^{cc^\prime}_s}^T C \gamma_5 S^{bb^\prime}_d)
\nonumber \\ & &
-\beta \gamma_5 S^{aa^\prime}_s \gamma_5 C {S^{cc^\prime}_u}^T C S^{bb^\prime}_d
-\beta \gamma_5 S^{aa^\prime}_s \gamma_5 C {S^{cc^\prime}_d}^T C S^{bb^\prime}_u
+\beta \gamma_5 S^{aa^\prime}_s \mbox{Tr}(C {S^{cc^\prime}_d}^T C S^{bb^\prime}_u \gamma_5)
\nonumber \\ & &
-\beta \gamma_5 S^{aa^\prime}_d \gamma_5 C {S^{cc^\prime}_u}^T C S^{bb^\prime}_s
-\beta \gamma_5 S^{aa^\prime}_d \gamma_5 C {S^{cc^\prime}_s}^T C S^{bb^\prime}_u
+\beta \gamma_5 S^{aa^\prime}_d \mbox{Tr}(C {S^{cc^\prime}_s}^T C S^{bb^\prime}_u \gamma_5)
\nonumber \\ & &
-\beta \gamma_5 S^{aa^\prime}_u \gamma_5 C {S^{cc^\prime}_d}^T C S^{bb^\prime}_s
-\beta \gamma_5 S^{aa^\prime}_u \gamma_5 C {S^{cc^\prime}_s}^T C S^{bb^\prime}_d
+\beta \gamma_5 S^{aa^\prime}_u \mbox{Tr}(C {S^{cc^\prime}_s}^T C S^{bb^\prime}_d \gamma_5)
\nonumber \\ & &
-\beta^2 \gamma_5 S^{aa^\prime}_s C {S^{cc^\prime}_u}^T C S^{bb^\prime}_d \gamma_5
-\beta^2 \gamma_5 S^{aa^\prime}_s C {S^{cc^\prime}_d}^T C S^{bb^\prime}_u \gamma_5
+\beta^2 \gamma_5 S^{aa^\prime}_s \gamma_5 \mbox{Tr}(C {S^{cc^\prime}_d}^T C S^{bb^\prime}_u)
\nonumber \\ & &
-\beta^2 \gamma_5 S^{aa^\prime}_d C {S^{cc^\prime}_u}^T C  S^{bb^\prime}_s \gamma_5
-\beta^2 \gamma_5 S^{aa^\prime}_d C {S^{cc^\prime}_s}^T C S^{bb^\prime}_u \gamma_5
+\beta^2 \gamma_5 S^{aa^\prime}_d \gamma_5 \mbox{Tr}(C {S^{cc^\prime}_s}^T C S^{bb^\prime}_u)
\nonumber \\ & &
-\beta^2 \gamma_5 S^{aa^\prime}_u C {S^{cc^\prime}_d}^T C S^{bb^\prime}_s \gamma_5
-\beta^2 \gamma_5 S^{aa^\prime}_u C {S^{cc^\prime}_s}^T C S^{bb^\prime}_d \gamma_5
+\beta^2 \gamma_5 S^{aa^\prime}_u \gamma_5 \mbox{Tr}(C {S^{cc^\prime}_s}^T C S^{bb^\prime}_d)\;\},
\label{master11_lambda_s}
\end{eqnarray}
and for the common $\Lambda$:
\begin{eqnarray}
& &
\langle\Omega\,|\, T\{\;\eta^{\Lambda_c}_{\scriptscriptstyle 1/2}(x)\,
\bar{\eta}^{\Lambda_c}_{\scriptscriptstyle 1/2}(0)\;\}\,|\,\Omega\rangle
=-2\epsilon^{abc}\epsilon^{a^\prime b^\prime c^\prime} \{\;
\nonumber \\ & &
-S^{aa^\prime}_d \gamma_5 C {S^{cc^\prime}_s}^T C  \gamma_5 S^{bb^\prime}_u
+S^{aa^\prime}_d \mbox{Tr}(C {S^{cc^\prime}_s}^T C \gamma_5 S^{bb^\prime}_u \gamma_5)
\nonumber \\ & &
-S^{aa^\prime}_u \gamma_5 C {S^{cc^\prime}_s}^T C \gamma_5 S^{bb^\prime}_d
+S^{aa^\prime}_u \mbox{Tr}(C {S^{cc^\prime}_s}^T C \gamma_5 S^{bb^\prime}_d \gamma_5)
\nonumber \\ & &
-\beta S^{aa^\prime}_d C {S^{cc^\prime}_s}^T C  \gamma_5 S^{bb^\prime}_u \gamma_5 
+\beta S^{aa^\prime}_d \gamma_5 \mbox{Tr}(C {S^{cc^\prime}_s}^T C \gamma_5 S^{bb^\prime}_u)
\nonumber \\ & &
-\beta S^{aa^\prime}_u C {S^{cc^\prime}_s}^T C  \gamma_5 S^{bb^\prime}_d \gamma_5
+\beta S^{aa^\prime}_u \gamma_5 \mbox{Tr}(C {S^{cc^\prime}_s}^T C \gamma_5 S^{bb^\prime}_d)
\nonumber \\ & &
-\beta \gamma_5 S^{aa^\prime}_d \gamma_5 C {S^{cc^\prime}_s}^T C S^{bb^\prime}_u
+\beta \gamma_5 S^{aa^\prime}_d \mbox{Tr}(C {S^{cc^\prime}_s}^T C S^{bb^\prime}_u \gamma_5)
\nonumber \\ & &
-\beta \gamma_5 S^{aa^\prime}_u \gamma_5 C {S^{cc^\prime}_s}^T C S^{bb^\prime}_d
+\beta \gamma_5 S^{aa^\prime}_u \mbox{Tr}(C {S^{cc^\prime}_s}^T C S^{bb^\prime}_d \gamma_5)
\nonumber \\ & &
-\beta^2 \gamma_5 S^{aa^\prime}_d C {S^{cc^\prime}_s}^T C S^{bb^\prime}_u \gamma_5
+\beta^2 \gamma_5 S^{aa^\prime}_d \gamma_5 \mbox{Tr}(C {S^{cc^\prime}_s}^T C S^{bb^\prime}_u)
\nonumber \\ & &
-\beta^2 \gamma_5 S^{aa^\prime}_u C {S^{cc^\prime}_s}^T C S^{bb^\prime}_d \gamma_5
+\beta^2 \gamma_5 S^{aa^\prime}_u \gamma_5 \mbox{Tr}(C {S^{cc^\prime}_s}^T C S^{bb^\prime}_d)\;\}.
\label{master11_lambda_c}
\end{eqnarray}

\subsection{Mixed Correlators}

The mixed correlators come from the generalized spin-1/2 current
$\eta_{1/2, \mu} =\gamma_\mu\gamma_5\, \eta_{1/2}$ with $\eta_{1/2}$ as given 
in Eq.~(\ref{nucleon_beta}), and the spin-3/2 current $\bar{\eta}_{3/2, \nu}$ as given 
in Eq.~(\ref{nuc32}).
For the nucleon, the master formula is given by
\begin{eqnarray}
& &
\langle\Omega\,|\, T\{\;\eta^{N}_{\scriptscriptstyle \mu,1/2}(x)\,
\bar{\eta}^{N}_{\scriptscriptstyle \nu,3/2}(0)\;\}\,|\,\Omega\rangle
=-2\epsilon^{abc}\epsilon^{a^\prime b^\prime c^\prime}\{\;
\nonumber \\ & &
-2\gamma_\mu\gamma_5  S^{aa^\prime}_u \sigma_{\rho\lambda}
 C {S^{cc^\prime}_u}^T C \gamma_5 S^{bb^\prime}_d \gamma_\nu \sigma^{\rho\lambda}
-\gamma_\mu\gamma_5  S^{aa^\prime}_u \sigma_{\rho\lambda}
 C {S^{cc^\prime}_d}^T C \gamma_5 S^{bb^\prime}_u \gamma_\nu \sigma^{\rho\lambda}
\nonumber \\ & &
+\gamma_\mu\gamma_5  S^{aa^\prime}_u \gamma_\nu \sigma_{\rho\lambda}
 \mbox{Tr}(\gamma_5 S^{bb^\prime}_d \sigma^{\rho\lambda} C {S^{cc^\prime}_u}^T C)
-2\beta\gamma_\mu S^{aa^\prime}_u \sigma_{\rho\lambda}
 C {S^{cc^\prime}_u}^T C  S^{bb^\prime}_d \gamma_\nu \sigma^{\rho\lambda}
\nonumber \\ & &
-\beta\gamma_\mu  S^{aa^\prime}_u \sigma_{\rho\lambda}
 C {S^{cc^\prime}_d}^T C S^{bb^\prime}_u \gamma_\nu \sigma^{\rho\lambda}
+\beta \gamma_\mu S^{aa^\prime}_u \gamma_\nu \sigma_{\rho\lambda}
 \mbox{Tr}( S^{bb^\prime}_d \sigma^{\rho\lambda} C {S^{cc^\prime}_u}^T C)
\; \}.
\label{master12}
\end{eqnarray}
The master formulas for $\Sigma$ and  $\Xi$ are obtained in the same way as the spin-1/2 case 
described in the previous section.
For the octet $\Lambda$, we have 
\begin{eqnarray}
& &
\langle\Omega\,|\, T\{\;\eta^{\Lambda_o}_{\scriptscriptstyle \mu,1/2}(x)\,
\bar{\eta}^{\Lambda_o}_{\scriptscriptstyle \nu,3/2}(0)\;\}\,|\,\Omega\rangle
={1\over 3}\epsilon^{abc}\epsilon^{a^\prime b^\prime c^\prime}\{\;
\nonumber \\ & &
+2\gamma_\mu\gamma_5 S^{aa^\prime}_s \sigma_{\rho\lambda}
 C {S^{cc^\prime}_u}^T C \gamma_5 S^{bb^\prime}_d \gamma_\nu \sigma^{\rho\lambda}
+2\gamma_\mu\gamma_5 S^{aa^\prime}_s \sigma_{\rho\lambda}
 C {S^{cc^\prime}_d}^T C \gamma_5 S^{bb^\prime}_u \gamma_\nu \sigma^{\rho\lambda}
\nonumber \\ & &
+4\gamma_\mu\gamma_5  S^{aa^\prime}_s \gamma_\nu \sigma_{\rho\lambda}
 \mbox{Tr}(\gamma_5 S^{bb^\prime}_d \sigma^{\rho\lambda} C {S^{cc^\prime}_u}^T C)
\nonumber \\ & &
+2\gamma_\mu\gamma_5 S^{aa^\prime}_d \sigma_{\rho\lambda}
 C {S^{cc^\prime}_u}^T C \gamma_5 S^{bb^\prime}_s \gamma_\nu \sigma^{\rho\lambda}
+\gamma_\mu\gamma_5 S^{aa^\prime}_d \sigma_{\rho\lambda}
 C {S^{cc^\prime}_s}^T C \gamma_5 S^{bb^\prime}_u \gamma_\nu \sigma^{\rho\lambda}
\nonumber \\ & &
+\gamma_\mu\gamma_5  S^{aa^\prime}_d \gamma_\nu \sigma_{\rho\lambda}
 \mbox{Tr}(\gamma_5 S^{bb^\prime}_s \sigma^{\rho\lambda} C {S^{cc^\prime}_u}^T C)
\nonumber \\ & &
-2\gamma_\mu\gamma_5 S^{aa^\prime}_u \sigma_{\rho\lambda}
 C {S^{cc^\prime}_d}^T C \gamma_5 S^{bb^\prime}_s \gamma_\nu \sigma^{\rho\lambda}
+\gamma_\mu\gamma_5 S^{aa^\prime}_u \sigma_{\rho\lambda}
 C {S^{cc^\prime}_s}^T C \gamma_5 S^{bb^\prime}_d \gamma_\nu \sigma^{\rho\lambda}
\nonumber \\ & &
+\gamma_\mu\gamma_5 S^{aa^\prime}_u \gamma_\nu \sigma_{\rho\lambda}
 \mbox{Tr}(\gamma_5 S^{bb^\prime}_s \sigma^{\rho\lambda} C {S^{cc^\prime}_d}^T C)
\nonumber \\ & &
+2\beta\gamma_\mu S^{aa^\prime}_s \sigma_{\rho\lambda}
 C {S^{cc^\prime}_u}^T C S^{bb^\prime}_d \gamma_\nu \sigma^{\rho\lambda}
+2\beta\gamma_\mu S^{aa^\prime}_s \sigma_{\rho\lambda}
 C {S^{cc^\prime}_d}^T C S^{bb^\prime}_u \gamma_\nu \sigma^{\rho\lambda}
\nonumber \\ & &
+4\beta\gamma_\mu S^{aa^\prime}_s \gamma_\nu \sigma_{\rho\lambda}
 \mbox{Tr}(S^{bb^\prime}_d \sigma^{\rho\lambda} C {S^{cc^\prime}_u}^T C)
\nonumber \\ & &
+2\beta\gamma_\mu S^{aa^\prime}_d \sigma_{\rho\lambda}
 C {S^{cc^\prime}_u}^T C S^{bb^\prime}_s \gamma_\nu \sigma^{\rho\lambda}
+\beta\gamma_\mu S^{aa^\prime}_d \sigma_{\rho\lambda}
 C {S^{cc^\prime}_s}^T C S^{bb^\prime}_u \gamma_\nu \sigma^{\rho\lambda}
\nonumber \\ & &
+\beta\gamma_\mu S^{aa^\prime}_d \gamma_\nu \sigma_{\rho\lambda}
 \mbox{Tr}(S^{bb^\prime}_s \sigma^{\rho\lambda} C {S^{cc^\prime}_u}^T C)
\nonumber \\ & &
-2\beta\gamma_\mu S^{aa^\prime}_u \sigma_{\rho\lambda}
 C {S^{cc^\prime}_d}^T C S^{bb^\prime}_s \gamma_\nu \sigma^{\rho\lambda}
+\beta\gamma_\mu S^{aa^\prime}_u \sigma_{\rho\lambda}
 C {S^{cc^\prime}_s}^T C S^{bb^\prime}_d \gamma_\nu \sigma^{\rho\lambda}
\nonumber \\ & &
+\beta\gamma_\mu S^{aa^\prime}_u \gamma_\nu \sigma_{\rho\lambda}
 \mbox{Tr}(S^{bb^\prime}_s \sigma^{\rho\lambda} C {S^{cc^\prime}_d}^T C)
\; \}.
\label{master12_lambda_o}
\end{eqnarray}

For the singlet $\Lambda$, we have
\begin{eqnarray}
& &
\langle\Omega\,|\, T\{\;\eta^{\Lambda_s}_{\scriptscriptstyle \mu,1/2}(x)\,
\bar{\eta}^{\Lambda_s}_{\scriptscriptstyle \nu,3/2}(0)\;\}\,|\,\Omega\rangle
={2\over 3}\epsilon^{abc}\epsilon^{a^\prime b^\prime c^\prime}\{\;
\nonumber \\ & &
-\gamma_\mu\gamma_5 S^{aa^\prime}_s \sigma_{\rho\lambda}
 C {S^{cc^\prime}_u}^T C \gamma_5 S^{bb^\prime}_d \gamma_\nu \sigma^{\rho\lambda}
-\gamma_\mu\gamma_5 S^{aa^\prime}_s \sigma_{\rho\lambda}
 C {S^{cc^\prime}_d}^T C \gamma_5 S^{bb^\prime}_u \gamma_\nu \sigma^{\rho\lambda}
\nonumber \\ & &
+\gamma_\mu\gamma_5  S^{aa^\prime}_s \gamma_\nu \sigma_{\rho\lambda}
 \mbox{Tr}(\gamma_5 S^{bb^\prime}_d \sigma^{\rho\lambda} C {S^{cc^\prime}_u}^T C)
\nonumber \\ & &
-\gamma_\mu\gamma_5 S^{aa^\prime}_d \sigma_{\rho\lambda}
 C {S^{cc^\prime}_u}^T C \gamma_5 S^{bb^\prime}_s \gamma_\nu \sigma^{\rho\lambda}
+\gamma_\mu\gamma_5 S^{aa^\prime}_d \sigma_{\rho\lambda}
 C {S^{cc^\prime}_s}^T C \gamma_5 S^{bb^\prime}_u \gamma_\nu \sigma^{\rho\lambda}
\nonumber \\ & &
+\gamma_\mu\gamma_5  S^{aa^\prime}_d \gamma_\nu \sigma_{\rho\lambda}
 \mbox{Tr}(\gamma_5 S^{bb^\prime}_s \sigma^{\rho\lambda} C {S^{cc^\prime}_u}^T C)
\nonumber \\ & &
+\gamma_\mu\gamma_5 S^{aa^\prime}_u \sigma_{\rho\lambda}
 C {S^{cc^\prime}_d}^T C \gamma_5 S^{bb^\prime}_s \gamma_\nu \sigma^{\rho\lambda}
+\gamma_\mu\gamma_5 S^{aa^\prime}_u \sigma_{\rho\lambda}
 C {S^{cc^\prime}_s}^T C \gamma_5 S^{bb^\prime}_d \gamma_\nu \sigma^{\rho\lambda}
\nonumber \\ & &
+\gamma_\mu\gamma_5 S^{aa^\prime}_u \gamma_\nu \sigma_{\rho\lambda}
 \mbox{Tr}(\gamma_5 S^{bb^\prime}_s \sigma^{\rho\lambda} C {S^{cc^\prime}_d}^T C)
\nonumber \\ & &
-\beta\gamma_\mu S^{aa^\prime}_s \sigma_{\rho\lambda}
 C {S^{cc^\prime}_u}^T C S^{bb^\prime}_d \gamma_\nu \sigma^{\rho\lambda}
-\beta\gamma_\mu S^{aa^\prime}_s \sigma_{\rho\lambda}
 C {S^{cc^\prime}_d}^T C S^{bb^\prime}_u \gamma_\nu \sigma^{\rho\lambda}
\nonumber \\ & &
+\beta\gamma_\mu S^{aa^\prime}_s \gamma_\nu \sigma_{\rho\lambda}
 \mbox{Tr}(S^{bb^\prime}_d \sigma^{\rho\lambda} C {S^{cc^\prime}_u}^T C)
\nonumber \\ & &
-\beta\gamma_\mu S^{aa^\prime}_d \sigma_{\rho\lambda}
 C {S^{cc^\prime}_u}^T C S^{bb^\prime}_s \gamma_\nu \sigma^{\rho\lambda}
+\beta\gamma_\mu S^{aa^\prime}_d \sigma_{\rho\lambda}
 C {S^{cc^\prime}_s}^T C S^{bb^\prime}_u \gamma_\nu \sigma^{\rho\lambda}
\nonumber \\ & &
+\beta\gamma_\mu S^{aa^\prime}_d \gamma_\nu \sigma_{\rho\lambda}
 \mbox{Tr}(S^{bb^\prime}_s \sigma^{\rho\lambda} C {S^{cc^\prime}_u}^T C)
\nonumber \\ & &
+\beta\gamma_\mu S^{aa^\prime}_u \sigma_{\rho\lambda}
 C {S^{cc^\prime}_d}^T C S^{bb^\prime}_s \gamma_\nu \sigma^{\rho\lambda}
+\beta\gamma_\mu S^{aa^\prime}_u \sigma_{\rho\lambda}
 C {S^{cc^\prime}_s}^T C S^{bb^\prime}_d \gamma_\nu \sigma^{\rho\lambda}
\nonumber \\ & &
+\beta\gamma_\mu S^{aa^\prime}_u \gamma_\nu \sigma_{\rho\lambda}
 \mbox{Tr}(S^{bb^\prime}_s \sigma^{\rho\lambda} C {S^{cc^\prime}_d}^T C)
\; \}.
\label{master12_lambda_s}
\end{eqnarray}
For the common $\Lambda$, we have
\begin{eqnarray}
& &
\langle\Omega\,|\, T\{\;\eta^{\Lambda_c}_{\scriptscriptstyle \mu,1/2}(x)\,
\bar{\eta}^{\Lambda_c}_{\scriptscriptstyle \nu,3/2}(0)\;\}\,|\,\Omega\rangle
=\epsilon^{abc}\epsilon^{a^\prime b^\prime c^\prime}\{\;
\nonumber \\ & &
+\gamma_\mu\gamma_5 S^{aa^\prime}_d \sigma_{\rho\lambda}
 C {S^{cc^\prime}_s}^T C \gamma_5 S^{bb^\prime}_u \gamma_\nu \sigma^{\rho\lambda}
+\gamma_\mu\gamma_5  S^{aa^\prime}_d \gamma_\nu \sigma_{\rho\lambda}
 \mbox{Tr}(\gamma_5 S^{bb^\prime}_s \sigma^{\rho\lambda} C {S^{cc^\prime}_u}^T C)
\nonumber \\ & &
+\gamma_\mu\gamma_5 S^{aa^\prime}_u \sigma_{\rho\lambda}
 C {S^{cc^\prime}_s}^T C \gamma_5 S^{bb^\prime}_d \gamma_\nu \sigma^{\rho\lambda}
+\gamma_\mu\gamma_5 S^{aa^\prime}_u \gamma_\nu \sigma_{\rho\lambda}
\mbox{Tr}(\gamma_5 S^{bb^\prime}_s \sigma^{\rho\lambda}
C {S^{cc^\prime}_d}^T C)
\nonumber \\ & &
+\beta\gamma_\mu S^{aa^\prime}_d \sigma_{\rho\lambda}
C {S^{cc^\prime}_s}^T C S^{bb^\prime}_u
\gamma_\nu \sigma^{\rho\lambda}
+\beta\gamma_\mu S^{aa^\prime}_d \gamma_\nu \sigma_{\rho\lambda}
\mbox{Tr}(S^{bb^\prime}_s \sigma^{\rho\lambda}
C {S^{cc^\prime}_u}^T C)
\nonumber \\ & &
+\beta\gamma_\mu S^{aa^\prime}_u \sigma_{\rho\lambda}
C {S^{cc^\prime}_s}^T C S^{bb^\prime}_d
\gamma_\nu \sigma^{\rho\lambda}
+\beta\gamma_\mu S^{aa^\prime}_u \gamma_\nu \sigma_{\rho\lambda}
\mbox{Tr}(S^{bb^\prime}_s \sigma^{\rho\lambda}
C {S^{cc^\prime}_d}^T C)
\; \}.
\label{master12_lambda_c}
\end{eqnarray}

\subsection{Spin-3/2 Correlators}

Using the spin-3/2 interpolating field as given 
in Eq.~(\ref{nuc32}), the master formula is given for the nucleon:
\begin{eqnarray}
& &
\langle\Omega\,|\, T\{\;\eta^{N}_{\scriptscriptstyle \mu,3/2}(x)\,
\bar{\eta}^{N}_{\scriptscriptstyle \nu,3/2}(0)\;\}\,|\,\Omega\rangle
=\epsilon^{abc}\epsilon^{a^\prime b^\prime c^\prime}\,\sigma^{\rho\lambda} \gamma_\mu \{\; 
\nonumber \\ & &
-2S^{aa^\prime}_u \sigma_{\sigma\tau}
C {S^{cc^\prime}_u}^T C \sigma_{\rho\lambda} S^{bb^\prime}_d \gamma_\nu \sigma^{\sigma\tau}
+S^{aa^\prime}_u \sigma_{\sigma\tau}
C {S^{cc^\prime}_d}^T C \sigma_{\rho\lambda} S^{bb^\prime}_u \gamma_\nu \sigma^{\sigma\tau}
\nonumber \\ & &
+S^{aa^\prime}_u \gamma_\nu \sigma^{\sigma\tau}
\mbox{Tr}(S^{bb^\prime}_u \sigma_{\sigma\tau} C {S^{cc^\prime}_d}^T C \sigma_{\rho\lambda})
-2S^{aa^\prime}_d \sigma_{\sigma\tau}
C {S^{cc^\prime}_u}^T C \sigma_{\rho\lambda} S^{bb^\prime}_u \gamma_\nu \sigma^{\sigma\tau}
\nonumber \\ & &
+2S^{aa^\prime}_d \gamma_\nu \sigma^{\sigma\tau}
\mbox{Tr}(S^{bb^\prime}_u \sigma_{\sigma\tau} C {S^{cc^\prime}_u}^T C \sigma_{\rho\lambda})
\; \}.
\label{master22}
\end{eqnarray}
The master formulas for $\Sigma$ and $\Xi$ are obtained in the same way as described above.

For the octet $\Lambda$, we have
\begin{eqnarray}
& &
\langle\Omega\,|\, T\{\;\eta^{\Lambda_o}_{\scriptscriptstyle \mu,3/2}(x)\,
\bar{\eta}^{\Lambda_o}_{\scriptscriptstyle \nu,3/2}(0)\;\}\,|\,\Omega\rangle
={1\over 6}\epsilon^{abc}\epsilon^{a^\prime b^\prime c^\prime}\,\sigma^{\rho\lambda}
\gamma_\mu \{\;
\nonumber \\ & &
-2S^{aa^\prime}_s \sigma_{\sigma\tau}
 C {S^{cc^\prime}_d}^T C \sigma_{\rho\lambda} S^{bb^\prime}_u \gamma_\nu \sigma^{\sigma\tau}
+2S^{aa^\prime}_s \sigma_{\sigma\tau}
 C {S^{cc^\prime}_u}^T C \sigma_{\rho\lambda} S^{bb^\prime}_d \gamma_\nu \sigma^{\sigma\tau}
\nonumber \\ & &
+4S^{aa^\prime}_s \gamma_\nu \sigma^{\sigma\tau}
 \mbox{Tr}(S^{bb^\prime}_d \sigma_{\sigma\tau} C {S^{cc^\prime}_u}^T C \sigma_{\rho\lambda})
-S^{aa^\prime}_d \sigma_{\sigma\tau}
 C {S^{cc^\prime}_s}^T C \sigma_{\rho\lambda} S^{bb^\prime}_u \gamma_\nu \sigma^{\sigma\tau}
\nonumber \\ & &
+2S^{aa^\prime}_d \sigma_{\sigma\tau}
 C {S^{cc^\prime}_u}^T C \sigma_{\rho\lambda} S^{bb^\prime}_s \gamma_\nu \sigma^{\sigma\tau}
+S^{aa^\prime}_d \gamma_\nu \sigma^{\sigma\tau}
 \mbox{Tr}(S^{bb^\prime}_s \sigma_{\sigma\tau} C {S^{cc^\prime}_u}^T C \sigma_{\rho\lambda})
\nonumber \\ & &
-S^{aa^\prime}_u \sigma_{\sigma\tau}
 C {S^{cc^\prime}_s}^T C \sigma_{\rho\lambda} S^{bb^\prime}_d \gamma_\nu \sigma^{\sigma\tau}
-2S^{aa^\prime}_u \sigma_{\sigma\tau}
 C {S^{cc^\prime}_d}^T C \sigma_{\rho\lambda} S^{bb^\prime}_s \gamma_\nu \sigma^{\sigma\tau}
\nonumber \\ & &
+S^{aa^\prime}_u \gamma_\nu \sigma^{\sigma\tau}
 \mbox{Tr}(S^{bb^\prime}_s \sigma_{\sigma\tau} C {S^{cc^\prime}_d}^T C \sigma_{\rho\lambda})
\; \}.
\label{master22_lambda_o}
\end{eqnarray}
For the singlet $\Lambda$, we have
\begin{eqnarray}
& &
\langle\Omega\,|\, T\{\;\eta^{\Lambda_s}_{\scriptscriptstyle \mu,3/2}(x)\,
\bar{\eta}^{\Lambda_s}_{\scriptscriptstyle \nu,3/2}(0)\;\}\,|\,\Omega\rangle
={1\over 3}\epsilon^{abc}\epsilon^{a^\prime b^\prime c^\prime}\,\sigma^{\rho\lambda}
\gamma_\mu \{\;
\nonumber \\ & &
+S^{aa^\prime}_s \sigma_{\sigma\tau}
 C {S^{cc^\prime}_d}^T C \sigma_{\rho\lambda} S^{bb^\prime}_u \gamma_\nu \sigma^{\sigma\tau}
-S^{aa^\prime}_s \sigma_{\sigma\tau}
 C {S^{cc^\prime}_u}^T C \sigma_{\rho\lambda} S^{bb^\prime}_d \gamma_\nu \sigma^{\sigma\tau}
\nonumber \\ & &
+S^{aa^\prime}_s \gamma_\nu \sigma^{\sigma\tau}
 \mbox{Tr}(S^{bb^\prime}_d \sigma_{\sigma\tau} C {S^{cc^\prime}_u}^T C \sigma_{\rho\lambda})
-S^{aa^\prime}_d \sigma_{\sigma\tau}
 C {S^{cc^\prime}_s}^T C \sigma_{\rho\lambda} S^{bb^\prime}_u \gamma_\nu \sigma^{\sigma\tau}
\nonumber \\ & &
-S^{aa^\prime}_d \sigma_{\sigma\tau}
 C {S^{cc^\prime}_u}^T C \sigma_{\rho\lambda} S^{bb^\prime}_s \gamma_\nu \sigma^{\sigma\tau}
+S^{aa^\prime}_d \gamma_\nu \sigma^{\sigma\tau}
 \mbox{Tr}(S^{bb^\prime}_s \sigma_{\sigma\tau} C {S^{cc^\prime}_u}^T C \sigma_{\rho\lambda})
\nonumber \\ & &
-S^{aa^\prime}_u \sigma_{\sigma\tau}
 C {S^{cc^\prime}_s}^T C \sigma_{\rho\lambda} S^{bb^\prime}_d \gamma_\nu \sigma^{\sigma\tau}
+S^{aa^\prime}_u \sigma_{\sigma\tau}
 C {S^{cc^\prime}_d}^T C \sigma_{\rho\lambda} S^{bb^\prime}_s \gamma_\nu \sigma^{\sigma\tau}
\nonumber \\ & &
+S^{aa^\prime}_u \gamma_\nu \sigma^{\sigma\tau}
 \mbox{Tr}(S^{bb^\prime}_s \sigma_{\sigma\tau} C {S^{cc^\prime}_d}^T C \sigma_{\rho\lambda})
\; \}.
\label{master22_lambda_s}
\end{eqnarray}
For the common $\Lambda$,
\begin{eqnarray}
& &
\langle\Omega\,|\, T\{\;\eta^{\Lambda_c}_{\scriptscriptstyle \mu,3/2}(x)\,
\bar{\eta}^{\Lambda_c}_{\scriptscriptstyle \nu,3/2}(0)\;\}\,|\,\Omega\rangle
={1\over 2}\epsilon^{abc}\epsilon^{a^\prime b^\prime c^\prime}\,\sigma^{\rho\lambda}
\gamma_\mu \{\;
\nonumber \\ & &
-S^{aa^\prime}_d \sigma_{\sigma\tau}
 C {S^{cc^\prime}_s}^T C \sigma_{\rho\lambda} S^{bb^\prime}_u \gamma_\nu \sigma^{\sigma\tau}
+S^{aa^\prime}_d \gamma_\nu \sigma^{\sigma\tau}
 \mbox{Tr}(S^{bb^\prime}_s \sigma_{\sigma\tau} C {S^{cc^\prime}_u}^T C \sigma_{\rho\lambda})
\nonumber \\ & &
-S^{aa^\prime}_u \sigma_{\sigma\tau}
 C {S^{cc^\prime}_s}^T C \sigma_{\rho\lambda} S^{bb^\prime}_d \gamma_\nu \sigma^{\sigma\tau}
+S^{aa^\prime}_u \gamma_\nu \sigma^{\sigma\tau}
 \mbox{Tr}(S^{bb^\prime}_s \sigma_{\sigma\tau} C {S^{cc^\prime}_d}^T C \sigma_{\rho\lambda})
\; \}.
\label{master22_lambda_c}
\end{eqnarray}

\end{document}